\documentclass[aps,twocolumn,preprintnumbers,nofootinbib,superscriptaddress]{revtex4}
\usepackage{eurosym}
\usepackage{amsmath}
\usepackage{graphicx}
\usepackage{amsfonts}
\usepackage{array}
\usepackage{amsthm}
\usepackage{bm}
\usepackage{palatino}
\usepackage{mathpazo}
\usepackage{supertabular}
\usepackage{subfig}

\usepackage[colorlinks]{hyperref}
\usepackage{subfig}
\usepackage[font=small,labelfont=bf,justification=justified,format=plain]{caption}
\usepackage[font=small,labelfont=bf,justification=justified]{caption}
\usepackage{color}
\usepackage{booktabs}
\usepackage{multirow}
\usepackage[labelfont=bf,labelsep=quad,justification=justified]{caption}
\usepackage{xcolor}
\newcommand{\be}{\begin{equation}}
\newcommand{\ee}{\end{equation}}
\newcommand{\ba}{\begin{eqnarray}}
\newcommand{\ea}{\end{eqnarray}}
\newcommand{\bal}{\begin{align}}
\newcommand{\eal}{\end{align}}
\newcommand{\lb}{\label}

\newcommand{\bw}{\begin{widetext}}
\newcommand{\ew}{\end{widetext}}

\begin{document}
\title{Shadows of Sgr A$^{*}$ Black Hole Surrounded by Superfluid Dark Matter Halo}
\author{Kimet Jusufi}
\email{kimet.jusufi@unite.edu.mk}
\affiliation{Physics Department, State University of Tetovo, Ilinden Street nn, 1200,
Tetovo, North Macedonia}
\affiliation{Institute of Physics, Faculty of Natural Sciences and Mathematics, Ss. Cyril
and Methodius University, Arhimedova 3, 1000 Skopje, North Macedonia}
\author{Mubasher Jamil}
\email{mjamil@zjut.edu.cn(corresponding author)}
\affiliation{Institute for Theoretical Physics and Cosmology, Zhejiang University of Technology, Hangzhou, 310023, China}
\affiliation{Department of Mathematics, School of Natural
	Sciences (SNS), National University of Sciences and Technology
	(NUST), H-12, Islamabad, 44000, Pakistan}
\affiliation{United Center of Gravitational Wave Physics (UCGWP), Zhejiang University of Technology, Hangzhou, 310023, China}
\author{Tao Zhu}
\email{zhut05@zjut.edu.cn}
\affiliation{Institute for Theoretical Physics and Cosmology, Zhejiang University of Technology, Hangzhou, 310023, China}
\affiliation{United Center of Gravitational Wave Physics (UCGWP), Zhejiang University of Technology, Hangzhou, 310023, China}



\begin{abstract}
In this paper we construct a black hole solution surrounded by superfluid dark matter (BH-SFDM) and baryonic matter, and study their effects on the shadow images of the Sgr A$^{*}$ black hole. To achieve this goal, we have considered two density profiles for the baryonic matter described by the spherical exponential profile and the power law profile including a special case describing a totally dominated dark matter galaxy (TDDMG). Using the present values for the parameters of the superfluid dark matter and baryonic density profiles for the Sgr A$^{*}$ black hole, we find that the effects of the superfluid dark matter and baryonic matter on the size of shadows are almost negligible compared to the Kerr vacuum black hole. In addition, we find that by increasing the baryonic mass the shadow size increases considerably. This result can be linked to the matter distribution in the galaxy, namely the baryonic matter is mostly located in the galactic center and therefore, by increasing the baryonic matter can affect the size of black hole shadow compared to the totally dominated dark matter galaxy where we observe an increase of the angular diameter of  the Sgr A$^{*}$ black hole of the magnitude $10^{-5} \mu$arcsec.
\end{abstract}
\maketitle

\section{Introduction}

It is widely believed that the region at the centre of many galaxies contain black holes (BHs).  They are astrophysical objects which perform manifestations of extremely strong gravity such as formation of gigantic jets of particles and disruption of neighboring stars. From theoretical perspective, BHs serve as a perfect lab to test gravity in strong field regime. Recently, the Event Horizon Telescope (EHT) Collaboration announced their first results concerning the detection of an event horizon of a supermassive black hole at the center of a neighboring elliptical M87 galaxy \citep{m87}. Due to the gravitational lensing effect, the bright structure surrounding the black hole also known as the accretion disk appears distorted. Moreover, the region of accretion disk behind the black hole also gets visible due to the bending of light by black hole. The shadow image captured by EHT is in good agreement with the predictions of the spacetime geometry of black hole described by the Kerr metric. The studies of the shadow images of various black holes, together with the current and future observations, are expected to provide an important approach to understand the geometric structure of black holes or small deviations from Kerr metric in the strong field regime.

On the other hand, it is believed that up to $90\%$ of matter in the host galaxy of a central black hole  consists of dark matter. Thus, it is natural to expect that the dark matter halo which surrounds the central black hole could lead to small deviations from Kerr metric near the black hole horizon. With this motivation, the black hole solutions with dark matter halo and their effects on the black hole shadow has been studied in  \citep{Jusufi:2019nrn,Haroon:2018ryd,Haroon:2019new,Xu:2018wow,Hou:2018bar,Xu:2018mkl}.  In this paper, we consider a scenario of a BH solution surrounded by a halo, which contains both the dark matter and baryonic matter. For dark matter, we assume a superfluid dark matter model recently proposed in \cite{Berezhiani:2015bqa} and construct the corresponding black hole solution with the halo of dark matter and baryonic matter. 
As a specific example we are going to elaborate the shadow images of Sgr A$^{*}$ to estimate its angular diameter with the effect of the halo.

One aim here is to study the shadows of rotating black holes surrounded by SFDM and baryonic matter. As we mentioned, shadows possess interesting observational signatures of the black hole spacetime in the strong gravity regime and, in the future we expect that shadow observations can impose constrains on different gravity theories. It is interesting to mention that the characteristic map of a shadow image depends on the details of the surrounding environment around the black hole, while the shadow contour is determined only by the spacetime metric itself. In light of this, there have been extensive efforts to investigate shadows cast by different black hole and compact object spacetimes. The shadow of a Schwarzschild black hole was first studied in \cite{Synge66} and later in \cite{Luminet79} and the same for Kerr black hole was studied by \cite{DeWitt73}. Since then various authors have studied shadows of black holes in modified theories of gravity and wormholes geometries \cite{Zakharov05,Stuchlik:2019uvf,Shipley:2016omi,Gott:2018ocn,Takahashi:2005hy,Guo:2018kis,Mureika:2016efo,Moffat:2015kva,Hioki:2008zw,Li:2013jra,Abdujabbarov:2016hnw,Amir:2016cen,Saha:2018zas,Abdujabbarov:2012bn,Ayzenberg:2018jip,Cunha:2016wzk,Atamurotov:2013dpa,Atamurotov:2013sca,Bambi:2019tjh,Vagnozzi:2019apd,Zhu:2019ura,Amir:2018pcu,Jusufi:2019caq,Shaikh:2018kfv,Shaikh:2018lcc,Gyulchev:2019osj,Gyulchev:2018fmd,Abdujabbarov:2015pqp,Bambi:2008jg,Bambi:2010hf,Abdikamalov19, Kumar, 1,2,4,5,6,7,8,9}.

The plan of the paper is as follows: In Sec. II, we use the superfluid density profile to obtain the radial function of a spherically symmetric spacetime. In Sec. III, we study the effect of baryonic and SFDM in the spacetime metric, using the exponential and power law profile for the baryonic matter. In Sec. IV, we find a spherically symmetric black hole metric surrounded by SFDM. And then in Sec. V, we first construct the spinning black hole surrounded by dark matter halo based the spherically symmetric spacetime we obtained by applying the Newman-Janis method. With this spinning black hole,  we then study null geodesics and circular orbits to explore the shadows of Sgr A$^{*}$ black hole. In Sec. VI we present the main conclusions of this paper. We shall use the natural units $G=c=\hbar=1$ throughout the paper.

\section{Superfluid dark matter and spherically symmetric dark matter spacetime}
\subsection{Superfluid dark matter}

From the view of field theory, one can describe the superfluid by the theory of a spontaneously broken global $U(1)$ symmetry, in a state of finite $U(1)$ charge density. In particular, at low energy regime the relevant degree of freedom is the Goldstone boson for the broken symmetry of the phonon field $\psi$. On the other hand, the $U(1)$ symmetry acts non-linearly on $\psi$ as a shift symmetry, $\psi \rightarrow \psi + c$. In the non-relativistic regime, at finite chemical potential $\mu$, the most general effective theory is described by \citep{Berezhiani:2015bqa}
\begin{equation}
{\cal L}_{T=0} = P(X)\,,
\label{LT=0}
\end{equation}
in which $X=\mu-m\Phi +\dot{\psi}-(\vec{\nabla}\psi)^2/2m$. Note that $m$ is the particle mass and $\Phi$ represents the Newtonian gravitational potential. The main idea put forward in \cite{Berezhiani:2015bqa} is that the DM superfluid phonons are described by the modified Newtonian dynamics (MOND) type Lagrangian 
\begin{equation}
{\cal L}_{{\rm DM}, \,T =  0} = \frac{2\Lambda (2m)^{3/2}}{3} X\sqrt{|X|} \,.
\label{PMOND}
\end{equation}

At first sight the fractional power of $X$ may seem strange if~(\ref{PMOND}) represents for example a scalar field, however from the theory of phonons one can argue that exponent is crucial since it determines the superfluid equation of state. On the other hand, to obtain a mediate a force between baryons and the DM phonons one must have the coupling 
\begin{equation}
{\cal L}_{\rm int} = \alpha\Lambda \psi\, \rho_{B},
\label{coupling}
\end{equation}
where $\alpha$ is a constant while $\rho_{B}$ gives the baryonic density. Let us point out that at zero temperature, this superfluid theory has three parameters, namely the particle mass $m$, a self-interaction strength parameter $\Lambda$, and finally the coupling constant $\alpha$ between phonons and baryons. Starting from the action composed of the \eqref{PMOND} and~\eqref{coupling} one can obtain a MOND type force law. In particular the equation of motion for phonons is the following \cite{Berezhiani:2015bqa}
\begin{equation}
\vec{\nabla}\cdot \left( \frac{(\vec{\nabla}\psi)^2- 2m \hat{\mu}}{\sqrt{(\vec{\nabla}\psi)^2- 2m \hat{\mu}}}\vec{\nabla}\psi \right) =\frac{\alpha\rho_{\rm b}}{2}\,,
\end{equation}
with $\hat{\mu} \equiv \mu - m\Phi$. Using the limit $(\vec{\nabla}\psi)^2\gg 2m \hat{\mu}$ and ignoring the curl term one has
\begin{equation}
|\vec{\nabla}\psi| \vec{\nabla}\psi \simeq \alpha \, \vec{a}_{B}\,,
\label{gradphi}
\end{equation}
with $\vec{a}_{B}$ being the Newtonian acceleration due to baryons only. While the mediated acceleration resulting from \eqref{coupling} is 
\begin{equation}
\vec{a}_{phonon} = \alpha \Lambda\,\vec{\nabla}\psi\,. 
\label{aphigen}
\end{equation}
In this way it can be shown that
\begin{equation}
a_{phonon} = \sqrt{\alpha^3\Lambda^2 \,a_{B}}\,.
\label{deepMOND}
\end{equation}
From this result it is easy to see well known MOND acceleration by identifying
\begin{equation}
a_0 =\alpha^3\Lambda^2.
\label{a0us}
\end{equation}

From the galactic rotation curves for $a_0$ it is found

\begin{equation}
a_0^{\rm MOND} \simeq 1.2 \times 10^{-8}~{\rm cm}/{\rm s}^2\,.
\label{a0MOND}
\end{equation}

As we saw, the MOND type law obtained in the superfluid model is not exact value and only applies in the regime $(\vec{\nabla}\psi)^2\gg 2m \hat{\mu}$. More importantly, the total acceleration experienced by a test particle is a contribution of $\vec{a}_{B}$, $\vec{a}_{phonon}$ and $\vec{a}_{\rm DM}$.

\subsection{Spacetime metric in the presence of superfluid dark matter}

The density profile can be given in terms of dimensionless variables $\Xi$ and $\xi$, defined by \cite{Berezhiani:2015bqa}
\begin{equation}
\rho (r) =\rho_0\Xi^{1/2}, \,\,
r = \sqrt{\frac{\rho_0}{32\pi \Lambda^2 m^6}}~\xi,
\end{equation}
where $\rho_0 \equiv \rho(0)$ is the central density. From the
Lane-Emden equation:
\begin{equation}
\frac{1}{\xi^2}\frac{{\rm d}}{{\rm d}\xi} \left(\xi^2 \frac{{\rm d}\Xi}{{\rm d}\xi}\right) = - \Xi^{1/2}\,,
\label{LaneEmdenEqn}
\end{equation}
and using the boundary conditions $\Xi(0) = 1$ and $\Xi'(0) = 0$, the numerical solution yields
\begin{equation}
\xi_1 \simeq 2.75.
\end{equation}
The last equation determines the size of the condensate
\begin{equation}
R = \sqrt{\frac{\rho_0}{32\pi  \Lambda^2 m^6}}\; \xi_1 \,.
\label{Rhalo}
\end{equation}

A simple analytical form that provides a good fit is 
\begin{equation}
\Xi(\xi) = \cos\left(\frac{\pi}{2}\frac{\xi}{\xi_1}\right).
\end{equation}
The central density is related to the mass of the halo condensate as follows
\begin{equation}
\rho_0 =  \frac{M}{4\pi R^3} \frac{\xi_1}{|\Xi'(\xi_1)|} \,.
\end{equation}
From the numerics we find $\Xi'(\xi_1) \simeq -0.5$. Substituting~\eqref{Rhalo} in Eq. (15), we can solve for the central density
\begin{equation}
\rho_0 \simeq \left(\frac{M_{\rm DM}}{10^{12}M_\odot}\right)^{2/5} \left(\frac{m}{{\rm eV}}\right)^{18/5} \left(\frac{\Lambda}{{\rm meV}}\right)^{6/5}  \; 10^{-24}~{\rm g}/{\rm cm}^3\,.
\label{rhocentral}
\end{equation}
Meanwhile the halo radius is 
\begin{equation}
R \simeq \left(\frac{M_{\rm DM}}{10^{12}M_\odot}\right)^{1/5} \left(\frac{m}{{\rm eV}}\right)^{-6/5}\left(\frac{\Lambda}{{\rm meV}}\right)^{-2/5} \; 45~{\rm kpc}\,.
\label{halorad}
\end{equation}

Remarkably, for $m\sim {\rm eV}$ and $\Lambda\sim {\rm meV}$ we obtain DM halos of realistic size.  The mass profile of the dark matter
galactic halo is given by
\begin{equation}
M_{DM}\left( r\right) =4\pi \int_{0}^{r}\rho
_{DM}\left( r'\right) r^{'2}dr'.
\end{equation}

To solve the last integral we use Eqs. (10)-(14), and by rewriting the mass profile in terms of the new coordinate $\xi$. In particular we obtain
\begin{eqnarray}
      M_{DM}\left( r\right) 
      =\sin\left(\frac{\pi r}{2 R}\right)\frac{\pi r^2}{2 R}\frac{\rho_0^2}{8 \Lambda^2 m^6},\,\,\,\,\,r \leq R.
\end{eqnarray}

From the last equation one can find the tangential velocity $%
v_{tg}^{2}\left( r\right) =M_{DM}(r)/r$ of a test particle moving
in the dark halo in spherical symmetric space-time 
\begin{align}
      v^2_{tg}\left( r\right) =\sin(\frac{\pi r}{2 R})\frac{\pi r}{2 R}\frac{\rho_0^2}{8\Lambda^2 m^6}.
\end{align}

In this section, we derive the space-time geometry for pure dark matter. To do so, let us consider a static and spherically symmetric spacetime ansatz with pure dark matter in Schwarzschild coordinates can be written as follows 
\begin{equation}
\mathrm{d}s^{2}=-f(r)\mathrm{d}t^{2}+\frac{\mathrm{d}r^{2}}{g(r)}+r^{2}\left(
\mathrm{d}\theta ^{2}+\sin ^{2}\theta \mathrm{d}\phi ^{2}\right),  \label{5}
\end{equation}
in which $f(r)=g(r)$ are known as the redshift and shape
functions, respectively.
\begin{align}
      v_{tg}^{2}\left( r\right) = \frac{d\ln \sqrt{f(r)}}{d\ln r},
\end{align}
we find 
\begin{equation}
f(r)_{DM}=e^{-\frac{\rho_0^2 \cos(\frac{\pi \,r}{2 \,R})}{4 \Lambda^2 m^6 }}
\end{equation}
with the constraint 
\begin{equation}
\lim_{\rho_0 \to 0}\left( e^{-\frac{\rho_0^2 \cos(\frac{\pi \,r}{2 \,R})}{4 \Lambda^2 m^6 }} \right)=1
\end{equation}

In the standard cold dark matter picture, a halo of mass $M_{\rm DM} = 10^{12}\,M_\odot$. In addition, for $m=0.6$ eV and $\Lambda=0.2$ meV we find $\rho_0=0.02 \,\times 10^{-24} g/cm^3=9.2\,\times 10^{-8}eV^4$. Note that here we have used the conversion $10^{-19}$ g/cm$^{3} \simeq 0.4$ eV$^{4}$. 

\section{Effect of baryonic matter and superfluid dark matter }

In a realistic situation, galaxies consist of a  baryonic (normal) matter (consisting of stars
of mass $M_{star}$, ionized gas of mass $M_{gas}$, neutral hydrogen of mass $M_{HI}$ etc.,), the dark matter of mass $M_{DM}$, which we assume to be in the form of a superfluid. The total mass of the galaxy is therefore $M_B = M_{star} + M_{gas} + M_{HI} $ as the total baryonic
mass in the galaxy. But, as we saw by introducing the baryonic matter we must take into account a phonon-mediated force which describes the interaction between SFDM and baryonic matter. 

To calculate the radial acceleration on a test baryonic particle within the superfluid core this is given by the sum of three contributions \cite{Berezhiani:2017tth}
\begin{equation}
\vec{a}(r)=\vec{a}_B(r)+\vec{a}_{DM}(r)+\vec{a}_{phonon}(r),
\end{equation}
where the first term is just the baryonic Newtonian acceleration $a_{B} = M_{B}(r)/r^2$. The second term is the gravitational acceleration
from the superfluid core, $a_{DM} = M_{DM}(r)/r^2$. The third term is the phonon-mediated acceleration 
\begin{equation}
\vec{a}_{phonon}(r)=\alpha \Lambda\,\vec{\nabla}\psi.
\end{equation}

The strength of this term is set by $\alpha$ and $\Lambda$, or equivalently the critical acceleration
\begin{equation}
a_{phonon}(r)=\sqrt{a_0 a_B}, 
\end{equation}
which is nothing but the phonon force closely matches
the deep-MOND acceleration $a_0$ given by Eq.(8). Thus, in total, the centrifugal acceleration of the particle of a test particle moving in the dark halo in spherical symmetric space-time is given by
\begin{align}
      \frac{v^2_{tg}\left( r\right)}{r} =\frac{M_{DM}(r)}{r^2}+\frac{M_{BM}(r)}{r^2}+\sqrt{a_0 \frac{M_{BM}(r)}{r^2}}.
\end{align}

Outside the size of the superfluid ''core'' noted as $R$, there should be a different phase for DM matter, namely it is surrounded by DM particles in the normal phase, most likely described by a Navarro-Frenk-White (NFW) profile or some other effective type dark matter profile, most probably described by the empirical Burkert profile (see for example,  \cite{saluci1,Burkert,saluci2,Donato,lapi,saluci3,saluci4}). 

However, as was argued in \cite{Berezhiani:2017tth} we do not expect a sharp transition, but instead, there should be a transition region around at which the DM is nearly in equilibrium but is incapable of maintaining long range coherence. In the present work we are mostly interested to explore the effect of the SFDM core and, therefore, we are going to neglect the outer DM region in the normal phase given by the NFW profile. 

\subsection{Exponential profile }

As a toy model for describing the baryonic distribution, we consider a
spherical exponential profile of density given by \cite{Berezhiani:2017tth}
\begin{equation}
\rho_B(r)=\frac{M_B}{8 \pi L^3}e^{-\frac{r}{L}}
\end{equation}
where the characteristic scale $L$ plays the role of a radial
scale length and  $M_B$ is the baryonic mass. In that case the total mass can be written as a sum of masses of dark matter and the disc. The mass function of the stellar thin disk is written as
\begin{equation}
M_B(r)=\frac{M_B\left[2L^2-\left( 2 L^2+2r L+r^2 \right)e^{-\frac{r}{L}}\right]}{2 L^2}. 
\end{equation}

We observe from these profiles that by going in the outside region of core i.e. $r>>L$,  the mass $M_B(r)$ reduces to a constant $M_B$. In that way the last term in (28) becomes independent of $r$ and the empirical Tully--Fisher relation which describes the rotational curves of galaxies.  In our analyses, we shall simplify the calculations by considering a fixed baryonic mass $M_B$ in the last term in (28). Let us introduce the quantity
\begin{eqnarray}
v_0^2=\sqrt{M_B a_0},
\end{eqnarray}
and use the tangent velocity while assuming a spherically symmetric solution resulting with
\begin{eqnarray}
f(r)_{ BM+DM}& \approx & C\,r^{2 \,v_0^2}e^{-\frac{\rho_0^2 \cos(\frac{\pi \,r}{2 \,R})}{4 \Lambda^2 m^6 }}\\\notag
&\times &\exp\left[-\frac{ M_B}{r L}(2L-(2L+r)e^{-\frac{r}{L}})  \right],
\end{eqnarray}
valid for $r \leq R.$ In the present work, we shall focues on the HSB type galaxyes, such as the Milky Way galaxy. We can use $M_B=5-7 \times 10^{10} M{\odot}$ and $L= 2-3 $ kpc \cite{salucci}.

\subsection{Power law profile} 
As a second example, we assume the baryonic matter to be concentrated into an inner core of radius $r_c$, and that its mass profile $M_{B}(r)$ can be described by the simple relation \cite{Boehmer:2007um}
\begin{equation}
M_{B}(r)=M_B\left(\frac{r}{r+r_c}  \right)^{3\beta},
\end{equation}
in the present paper we shall be interested in the case $\beta = 1$, for high surface brightness galaxies (HSB).
Making use of the tangent velocity and assuming a spherically symmetric solution for the baryonic matter contribution in the case of HSB galaxies we find 
\begin{equation}
f(r)_{ BM+DM} \approx C\,\left(\frac{\pi r}{2 R}\right)^{2\,v_0^2}\,e^{-\frac{M_B(2r+r_c)}{(r+r_c)^2}}e^{-\frac{\rho_0^2 \cos(\frac{\pi \,r}{2 \,R})}{4 \Lambda^2 m^6 }}. 
\end{equation}
Note that the above analyses holds only inside the superfluid core with $r \leq R$.

  \begin{figure}
\includegraphics[width=6.1cm]{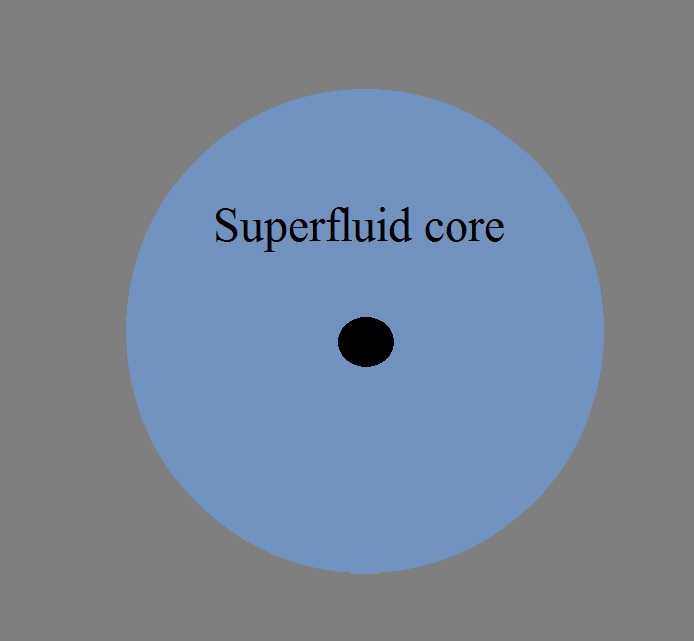}
  \caption{Schematic representation of the superfluid DM core and baryonic matter surrounding a black hole at the center.  Outside the superfluid core DM exists in a normal phase probably described by the NFW profile or Burkert profile. }
  \end{figure}
  
\subsection{Totally dominated dark matter galaxies}

There is one particular solution of interest which describes a totally dominated dark matter galaxies (TDDMG) or sometimes known as {\it dark galaxies}. To do this, we neglect the baryonic mass $M_{BM} \to 0$, and therefore $a_0\to 0$, yielding
\begin{equation}
f(r)_{TDDMG}=C\,e^{-\frac{\rho_0^2 \cos(\frac{\pi \,r}{2 \,R})}{4 \Lambda^2 m^6 }},
\end{equation}
valid for $r \leq R.$  This result is obtained directly from Eqs. (32) and (34) in the limit $M_B=0$. In other words this solution reduces to (23), as expected. Note that the integration constant $C$ can be absorbed in the time coordinate using $dt^2 \to C\, dt^2$.

\section{Black hole metric surrounded by baryonic matter and superfluid dark matter}

Let us now consider a more interesting scenario by adding a black hole present in the SFDM halo. To achieve this aim, we shall use the method introduced in \cite{Xu:2018wow}. Namely we need to compute the spacetime metric by assuming the Einstein field equations given by \cite{Xu:2018wow}
\begin{equation}
R^{\nu}_{~\mu}-\dfrac{1}{2}\delta^{\nu}_{~\mu}R=\kappa^{2}T^{\nu}_{~\mu}.
\label{SPBH7}
\end{equation}
where the corresponding energy-momentum tensors $T^{\nu}_{~\mu}=diag[-\rho,p_{r},p,p]$, where $\rho=\rho_{DM}+\rho_{BM}$, encodes the total contribution coming from the surrounded dark matter and baryonic matter, respectively.  The space-time metric including black hole is thus given by \cite{Xu:2018wow}
\begin{equation}
ds^{2}=-(f(r)+F_{1}(r))dt^{2}+\frac{dr^{2}}{g(r)+F_{2}(r)}+r^{2}d\Omega^2,
\label{SPBH9}
\end{equation}
where $d\Omega^2=d\theta^2+\sin^2\theta d\phi^2$ is the metric of a unit sphere. Moreover we have introduced the following quantities
\begin{equation}
\mathcal{F}(r)=f(r)+F_{1}(r),\,\,
\mathcal{G}(r)=g(r)+F_{2}(r).
\label{SPC10}
\end{equation}
In terms of these relations, the space-time metric describing a black hole in dark matter halo becomes \cite{Xu:2018wow}
\begin{equation}\notag
ds^{2}=-\exp[\int \dfrac{g(r)}{g(r)-\dfrac{2M}{r}}(\dfrac{1}{r}+\dfrac{f^{'}(r)}{f(r)})dr-\dfrac{1}{r} dr]dt^{2}
\end{equation}
\begin{equation}
+(g(r)-\dfrac{2M}{r})^{-1}dr^{2}
+r^{2}d\Omega^2.
\end{equation}

In the absence of the dark matter halo, i.e., $f(r)=g(r)=1$, the indefinite integral reduces to a constant \cite{Xu:2018wow}
\begin{eqnarray}\notag
F_{1}(r)+f(r)&=&\exp[\int \dfrac{g(r)}{g(r)+F_{2}(r)}(\dfrac{1}{r}+\dfrac{f^{'}(r)}{f(r)})-\dfrac{1}{r} dr]\\
&=& 1-\dfrac{2M}{r},
\end{eqnarray}
yielding $F_1(r)=-2M/r$. In that way one can obtain the black hole space-time metric with a surrounding matter 
\begin{eqnarray}
ds^{2}=-\mathcal{F}(r)dt^{2}+\frac{dr^2}{\mathcal{G}(r)}+r^2d\Omega^2.
\label{CDM11}
\end{eqnarray}

Thus the general black hole solution surrounded by SFDM with $\mathcal{F}(r)=\mathcal{G}(r)$ are given by
\begin{eqnarray}
\mathcal{F}(r)_{exp}&=&C\,r^{2\,v_0^2}\,e^{-\frac{\rho_0^2 \cos(\frac{\pi \,r}{2 \,R})}{4 \Lambda^2 m^6 }}\\\notag
&\times & \exp\left[-\frac{ M_B}{r L}(2L-(2L+r)e^{-\frac{r}{L}})   \right]-\dfrac{2M}{r},
\end{eqnarray}
and 
\begin{equation}
\mathcal{F}(r)_{power}=C\,\left(\frac{\pi r}{2 R}\right)^{2\,v_0^2}\,e^{-\frac{\rho_0^2 \cos(\frac{\pi \,r}{2 \,R})}{4 \Lambda^2 m^6 }}e^{-\frac{M_B(2r+r_c)}{(r+r_c)^2}}- \dfrac{2M}{r}.
\end{equation}

As a special case we obtain a black hole in a totally dominated dark matter galaxy given by
\begin{equation}
\mathcal{F}(r)_{TDDMG}=C\,e^{-\frac{\rho_0^2 \cos(\frac{\pi \,r}{2 \,R})}{4 \Lambda^2 m^6 }}-\frac{2M}{r},
\end{equation}
valid for $r \leq R.$  Note that $M$ is the black hole mass. In the limit $M=0$ our solution reduces to (23), as expected.

\section{Shadow of Sgr A$^{*}$ black hole surrounded by superfluid dark matter}

In this section we generalize the static and the spherical symmetric black hole solution to a spinning black hole surrounded by dark matter halo applying the Newman-Janis method and adapting the approach introduced in \cite{Azreg-Ainou:2014pra,Azreg-Ainou:2018tkl}. Applying the Newman-Janis method we obtain the spinning black hole metric surrounded by dark matter halo as follows (see Appendix A)
\begin{eqnarray}\notag
ds^2_{1,2}&=&-\left(1-\frac{2\Upsilon_{1,2}(r) r}{\Sigma}\right)dt^2+\frac{\Sigma}{\Delta_{1,2}}dr^2+\Sigma d\theta^2 \\\notag
&-& 2 a \sin^2\theta \frac{2\Upsilon_{1,2}(r) r}{\Sigma}dt d\phi \\
& +&\sin^2\theta \left[\frac{(r^2+a^2)^2-a^2\Delta_{1,2} \sin^2\theta}{\Sigma} \right] d\phi^2\label{metric}
\end{eqnarray}
where we have introduced
 \begin{equation}
\Upsilon_{1,2}(r)=\frac{r\, (1-\mathcal{F}_{1,2}(r))}{2},
\end{equation}
and identified $\mathcal{F}_{1}=\mathcal{F}(r)_{exp}$ and $\mathcal{F}_{2}=\mathcal{F}(r)_{power}$, respectively. As a special case we find the Kerr BH limit, $Y_{1,2}(r)=M$, if $a_0=M_B=\rho_0=0$.
 
 \begin{figure*}
\includegraphics[width=8.1cm]{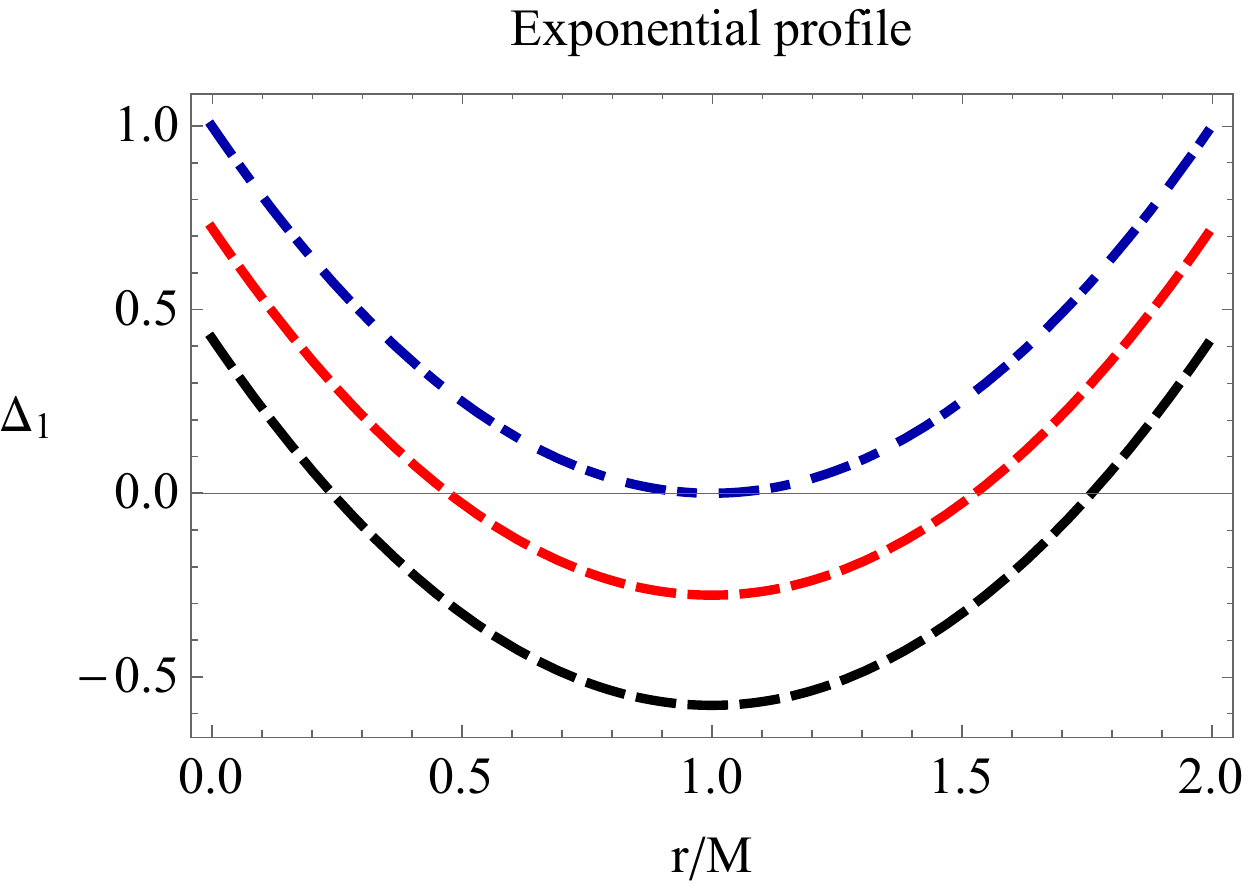}
\includegraphics[width=8.1cm]{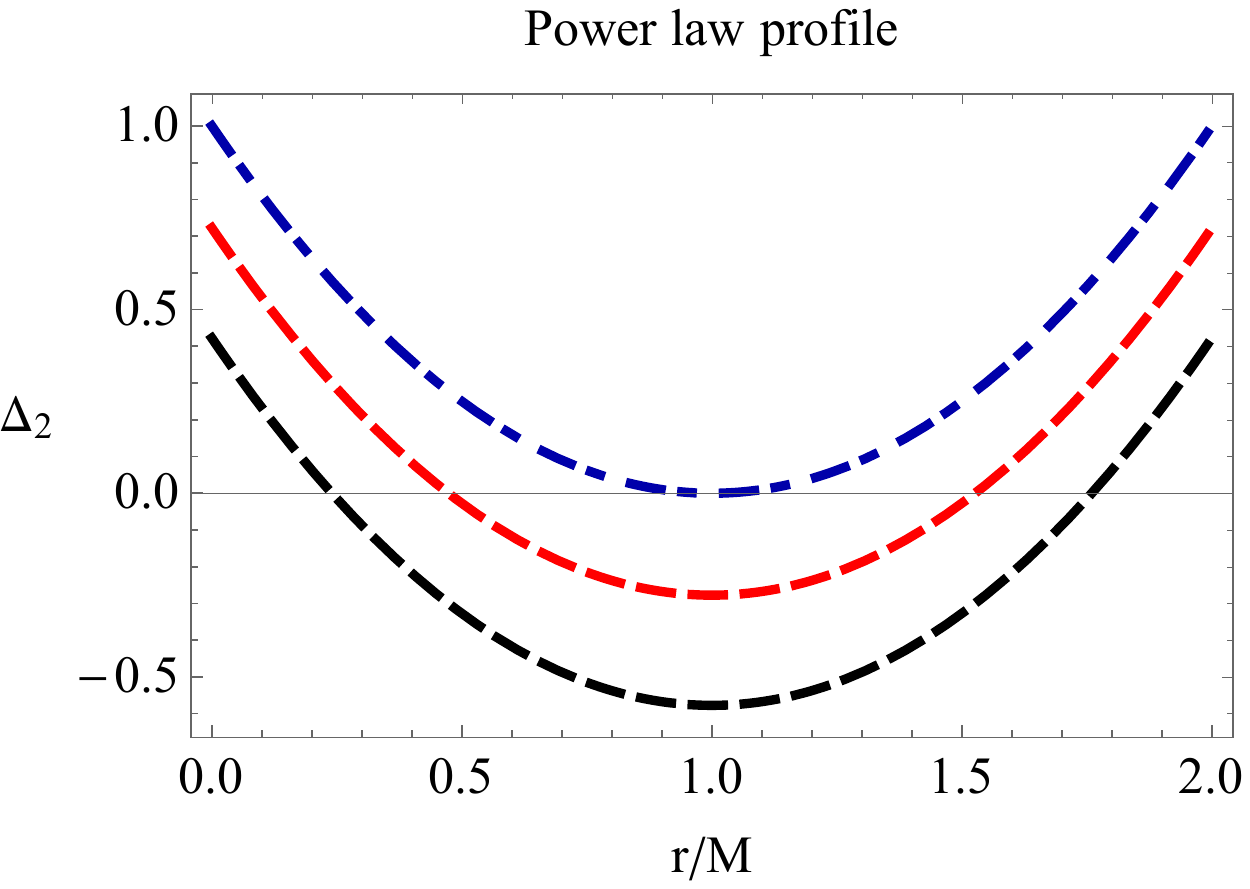}
  \caption{\label{figure2} Left panel: Variation of $\Delta_1$ as a function of $r$, for BH-SFDM using the spherical exponential profile. Right panel:  Variation of $\Delta_2$ as a function of $r$, for BH-SFDM using the power profile law. Note that $a=0.65$ (black curve), $a=0.85$ (red curve),  $a=1$ (blue curve), respectively. We have used  $m=0.6$ eV and $\Lambda=0.2$ meV we find $\rho_0=0.02 \,\times 10^{-24}$ g/cm$^3$, or in eV units, $\rho_0=9.2\,\times 10^{-8} $ eV$^4$. For the Milky Way galaxy we can use $M_B=6 \times 10^{10} M{\odot}=1.39\,\times 10^{4} M_{BH}$ and $r_c=2.6$ kpc$=1.24\,\times 10^{10}  M_{BH}$.}
  \end{figure*}
  
  \begin{figure*}
\includegraphics[width=8.1cm]{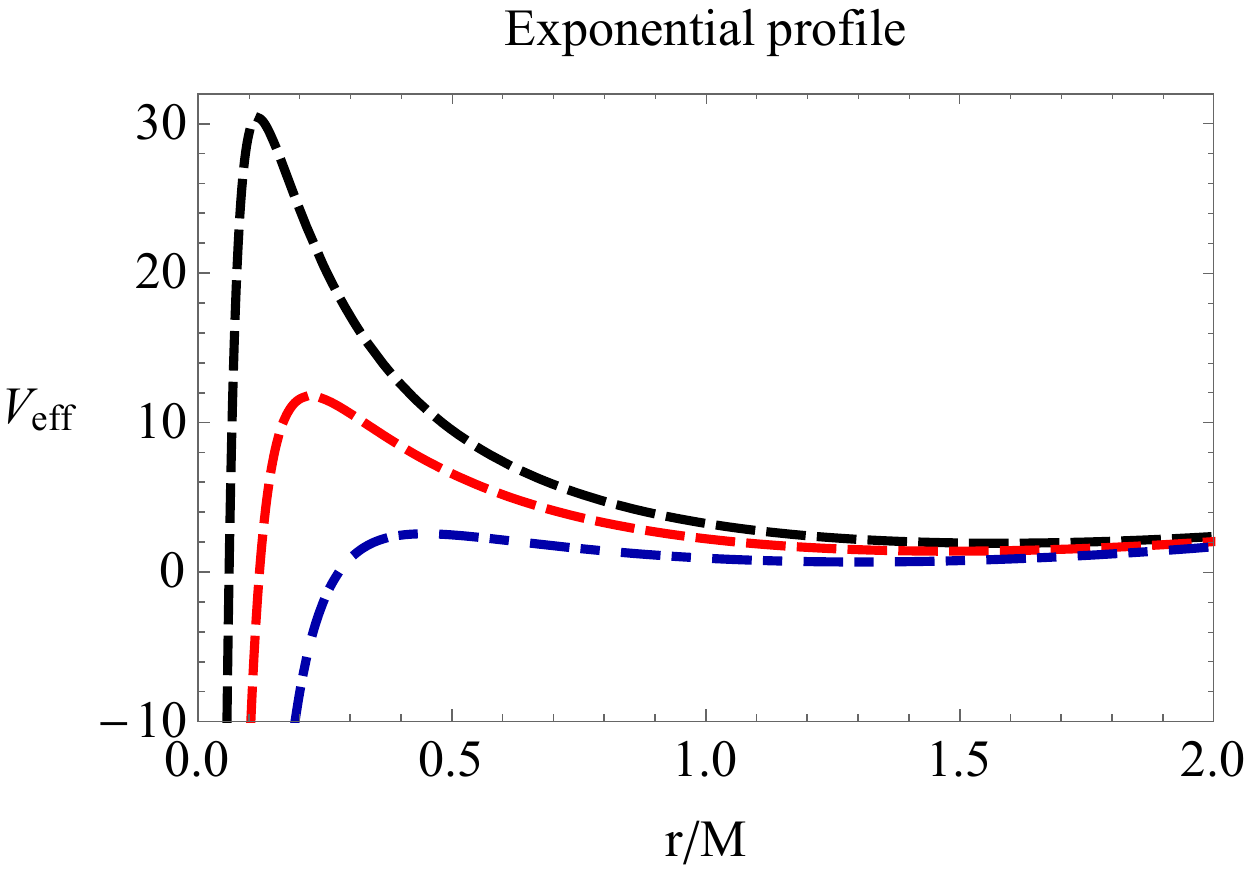}
\includegraphics[width=8.3cm]{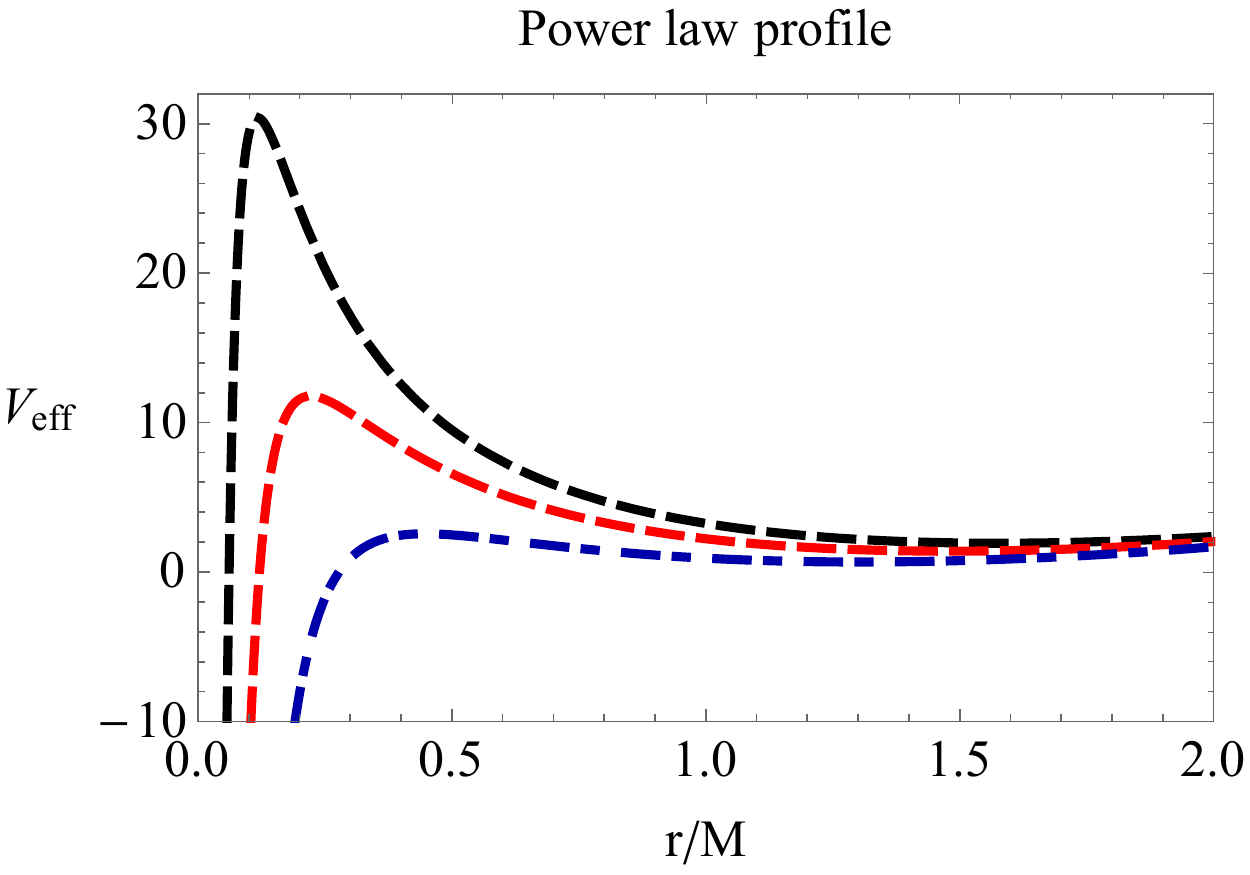}
\caption{ Left panel: The effective potential of photon moving for BH-SFDM using the exponsntial profile. Right panel: The effective potential of photon using thepower law profile. We have chosen $a=0.5$ (black color), $a=0.65$ (red color) and $a=0.85$ (blue color), respectively. } \label{B}
\end{figure*}

In addition we can explore the shape of the ergoregion of our black hole metric  \eqref{metric}, that is we can plot the shape of the ergoregion, say in the $xz$-plane. On the other hand the horizons of the black hole are found by solving $\Delta_{1,2}=0$, t.e., 
\begin{equation}
r^2\mathcal{F}(r)_{1,2}+a^2=0,
\end{equation}
while the inner and outer ergo-surfaces are obtained from solving $g_{tt}=0$, i.e., 
\begin{equation}
r^2\mathcal{F}(r)_{1,2}+a^2 \cos^2\theta=0.
\end{equation}

In Fig. 2 we plot $\Delta_{1,2}$ as a function of $r$ for different values of $a$. In the special case for some critical value $a=a_E$ (blue line) the two horizons coincide, in other words we have
an extremal black hole with degenerate horizons.  Beyond this critical value, $a>a_E$, there is no event horizon and the solution corresponds to a naked singularity.

\begin{figure*}
\includegraphics[width=7.1cm]{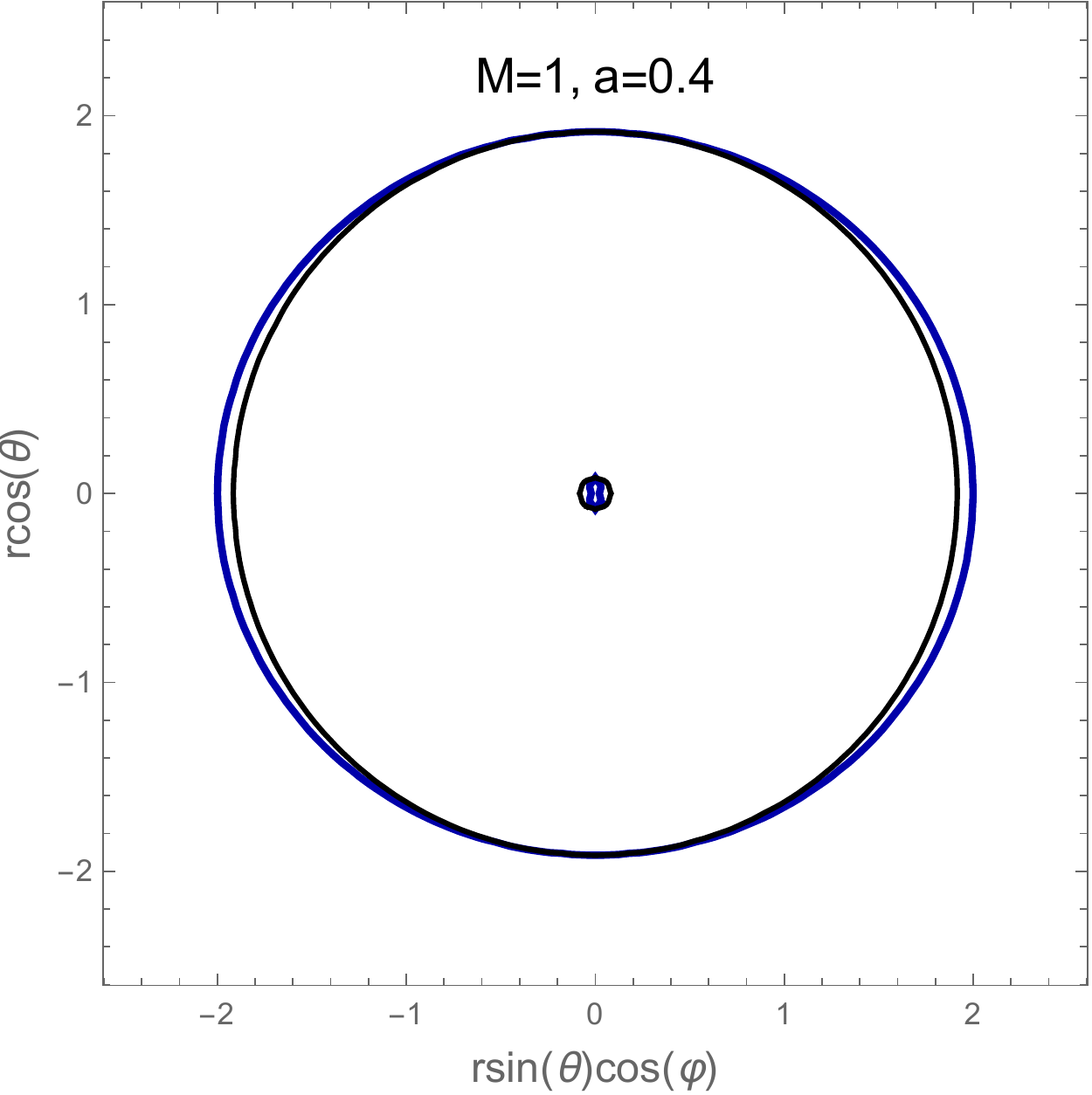}
\includegraphics[width=7.1cm]{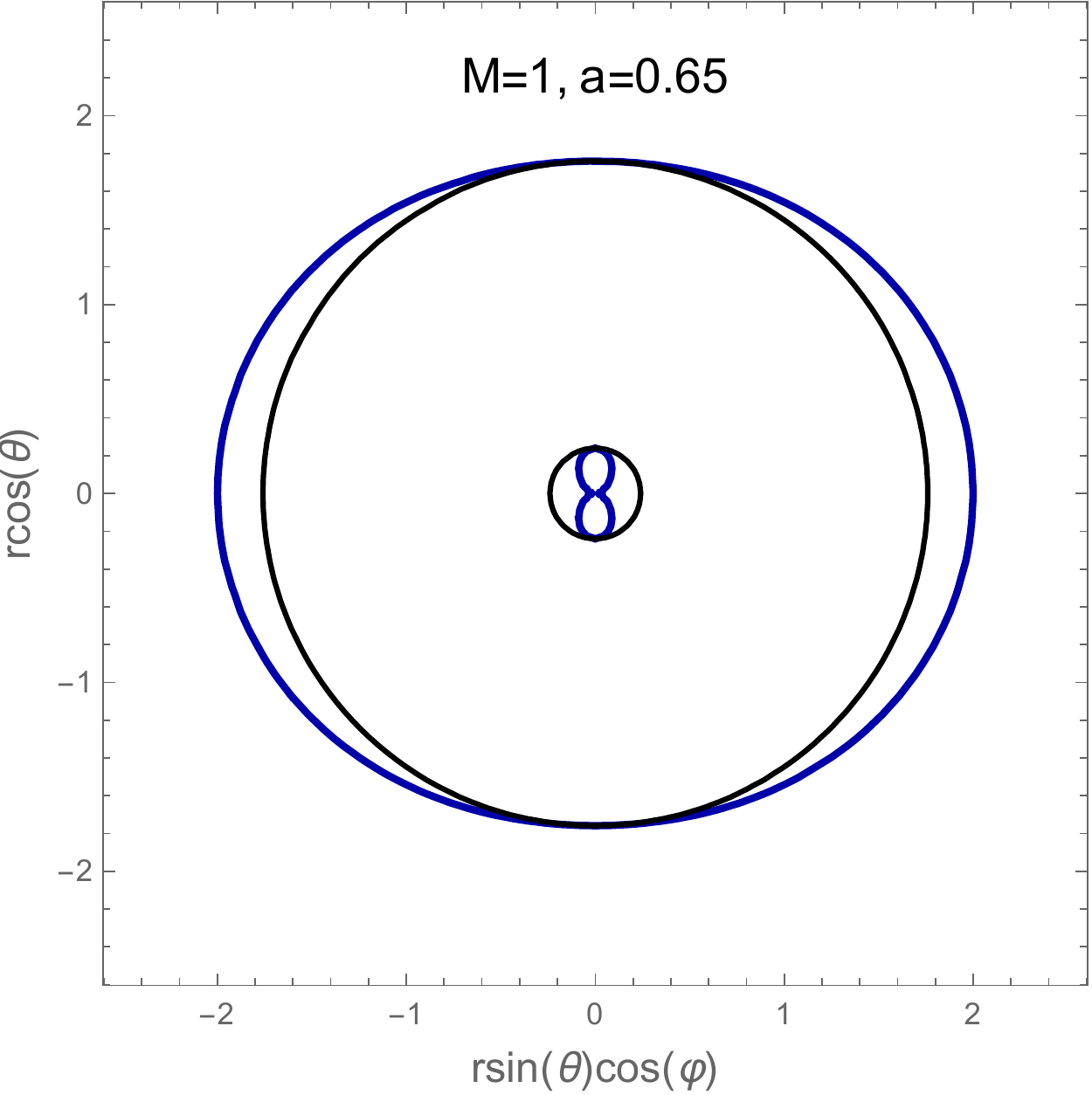}
\includegraphics[width=7.1cm]{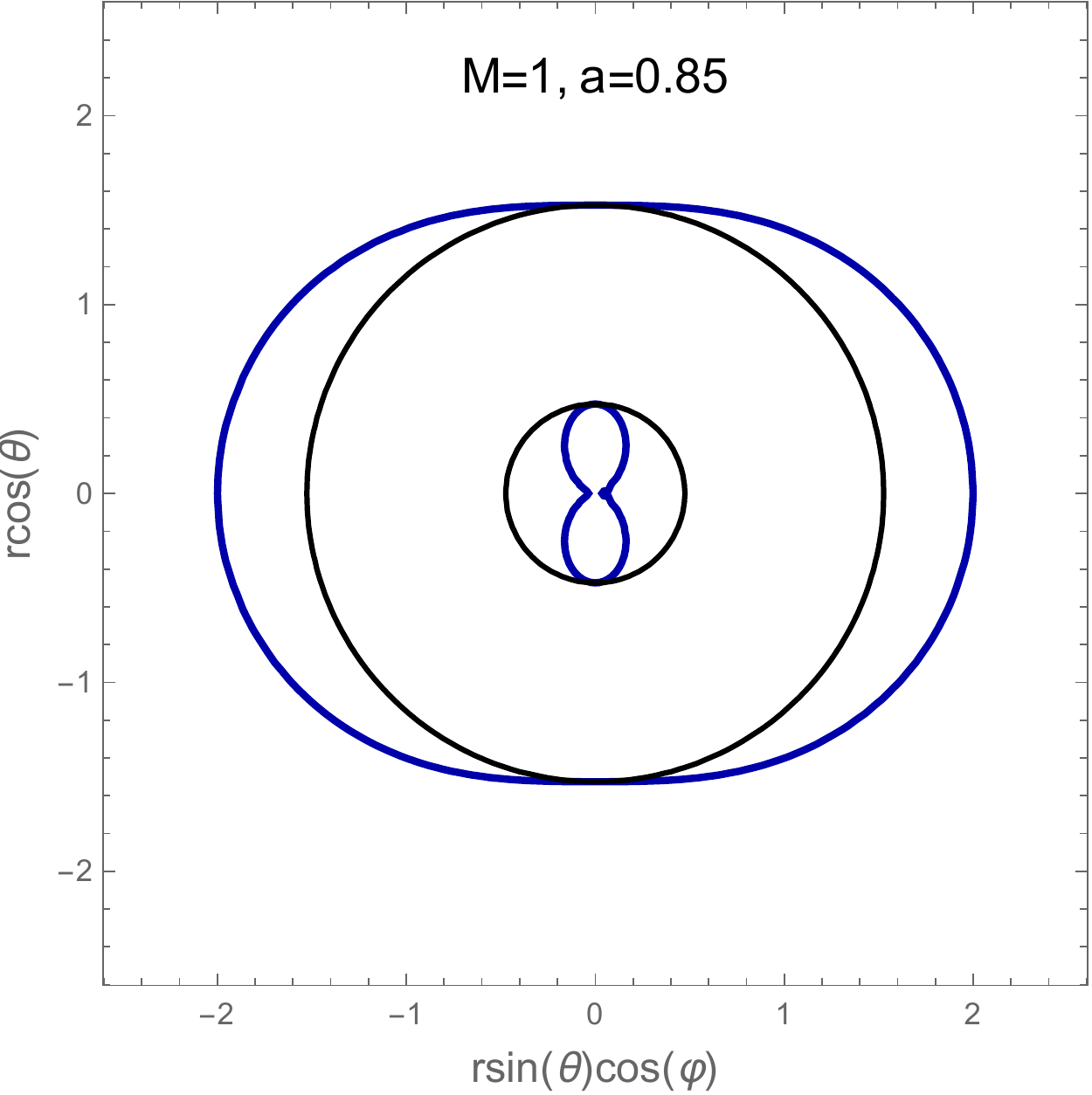}
\includegraphics[width=7.1cm]{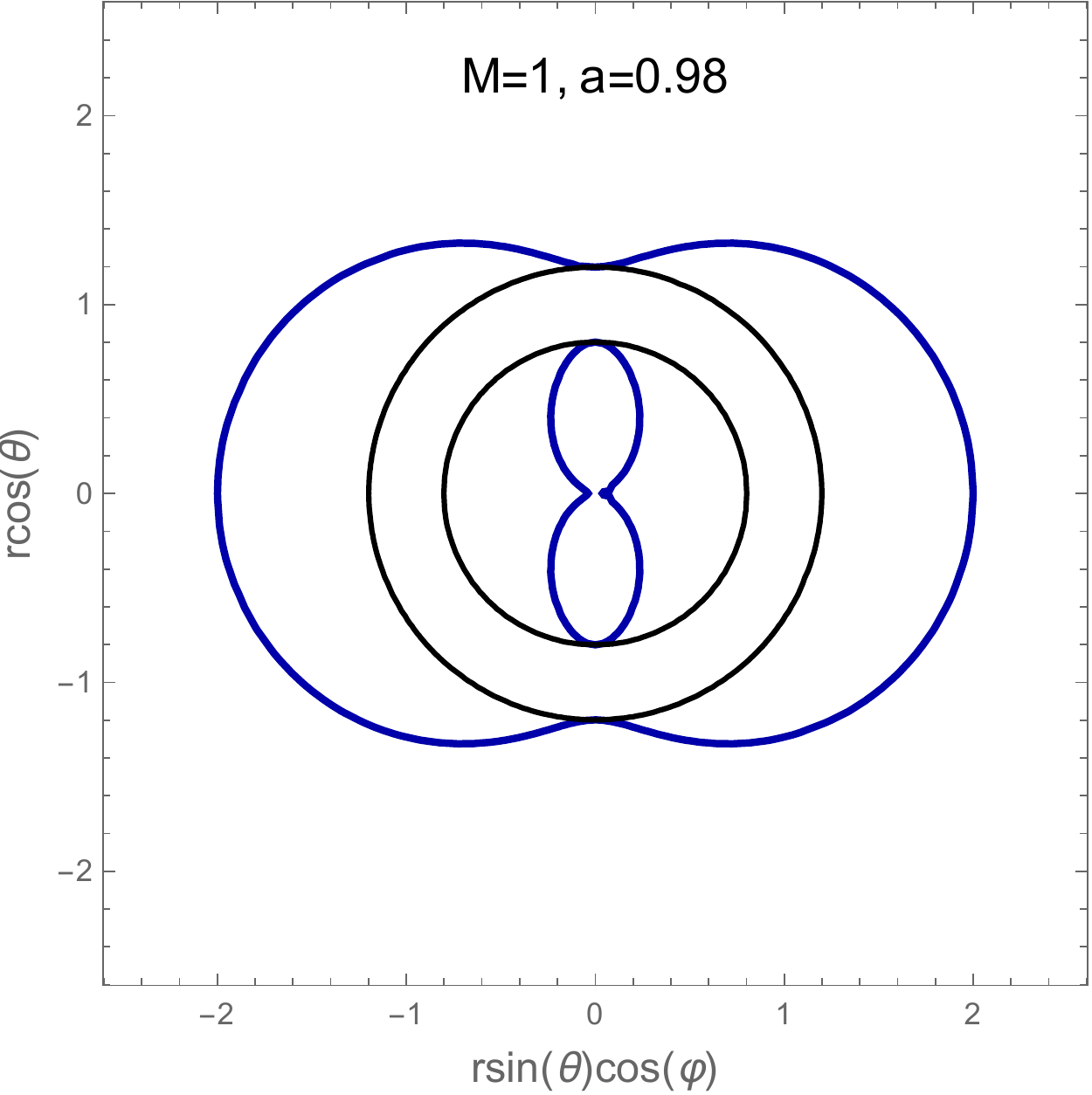}
\caption{The shape of the ergoregion and inner/outer horizons for different values of $a$ using the spherical exponential profile. We use $M=1$ in units of the Sgr A$^{*}$ black hole mass given by $M_{BH}=4.3\,\times 10^{6} M_{\odot}$ and $R=158$ kpc or $R=75.5\,\times 10^{10} M_{BH}$. Furthermore for $m=0.6$ eV and $\Lambda=0.2$ meV we find $\rho_0=0.02 \,\times 10^{-24}$ g/cm$^3$, or in eV units, $\rho_0=9.2\,\times 10^{-8} $ eV$^4$. For the normal matter contribution of the Milky Way galaxy we can use $M_B=6 \times 10^{10} M{\odot}=1.39\,\times 10^{4} M_{BH}$ and $L=2.6$ kpc$=1.24\,\times 10^{10}  M_{BH}$.  } \label{B}
\end{figure*}

\subsection{Geodesic equations}
In order to find the contour of a black hole shadow of the rotating spacetime (45), first we need to find the null geodesic equations using the Hamilton-Jacobi method given by
\begin{equation}
\frac{\partial \mathcal{S}}{\partial \tau}=-\frac{1}{2}g^{\mu\nu}\frac{\partial \mathcal{S}}{\partial x^\mu}\frac{\partial \mathcal{S}}{\partial x^\nu},
\label{eq:HJE}
\end{equation}
where $\tau$ is the affine parameter, and $\mathcal{S}$ is the Jacobi action. Due to the spectime symmetries there are two conserved quantities, namely the conserved energy $E=-p_t$ and the conserved angular momentum $L=p_\phi$, respectively. 

To find the separable solution of Eq. (\ref{eq:HJE}),  we need to express the action in the following form
\begin{equation}
\mathcal{S}=\frac{1}{2}\mu ^2 \tau - E t + J \phi + \mathcal{S}_{r}(r)+\mathcal{S}_{\theta}(\theta),
\label{eq:action_ansatz}
\end{equation}
in which $\mu$ gives the mass of the test particle. Of course, this gives $\mu=0$ in the case of the photon. 
That being said, it is straightforward to obtain the following equations of motions from the Hamilton-Jacobi equation
\begin{align}
\label{HJ3}
&\Sigma\frac{dt}{d\tau}=\frac{r^2+a^2}{\Delta_{1,2}}[E(r^2+a^2)-aL]-a(aE\sin^2\theta-L),\\
&\Sigma\frac{dr}{d\tau}=\sqrt{\mathcal{R}(r)},\\
&\Sigma\frac{d\theta}{d\tau}=\sqrt{\Theta(\theta)},\\
\label{HJ4}
&\Sigma\frac{d\phi}{d\tau}=\frac{a}{\Delta_{1,2}}[E(r^2+a^2)-aL]-\left(aE-\frac{L}{\sin^2\theta}\right),
\end{align}
where $\mathcal{R}(r)$ and $\Theta(\theta)$ are given by
\begin{align}
\label{HJ5}
&\mathcal{R}(r)=[E(r^2+a^2)-aL]^2-\Delta_{1,2}[m^2r^2+(aE-L)^2+\mathcal{K}],\\
&\Theta(\theta)=\mathcal{K}-\left(  \dfrac{L^2}{\sin^2\theta}-a^2E^2  \right) \cos^2\theta,
\end{align}
and $\mathcal{K}$ is the separation constant known as the Carter constant.

\begin{figure*}
\includegraphics[width=8.2cm]{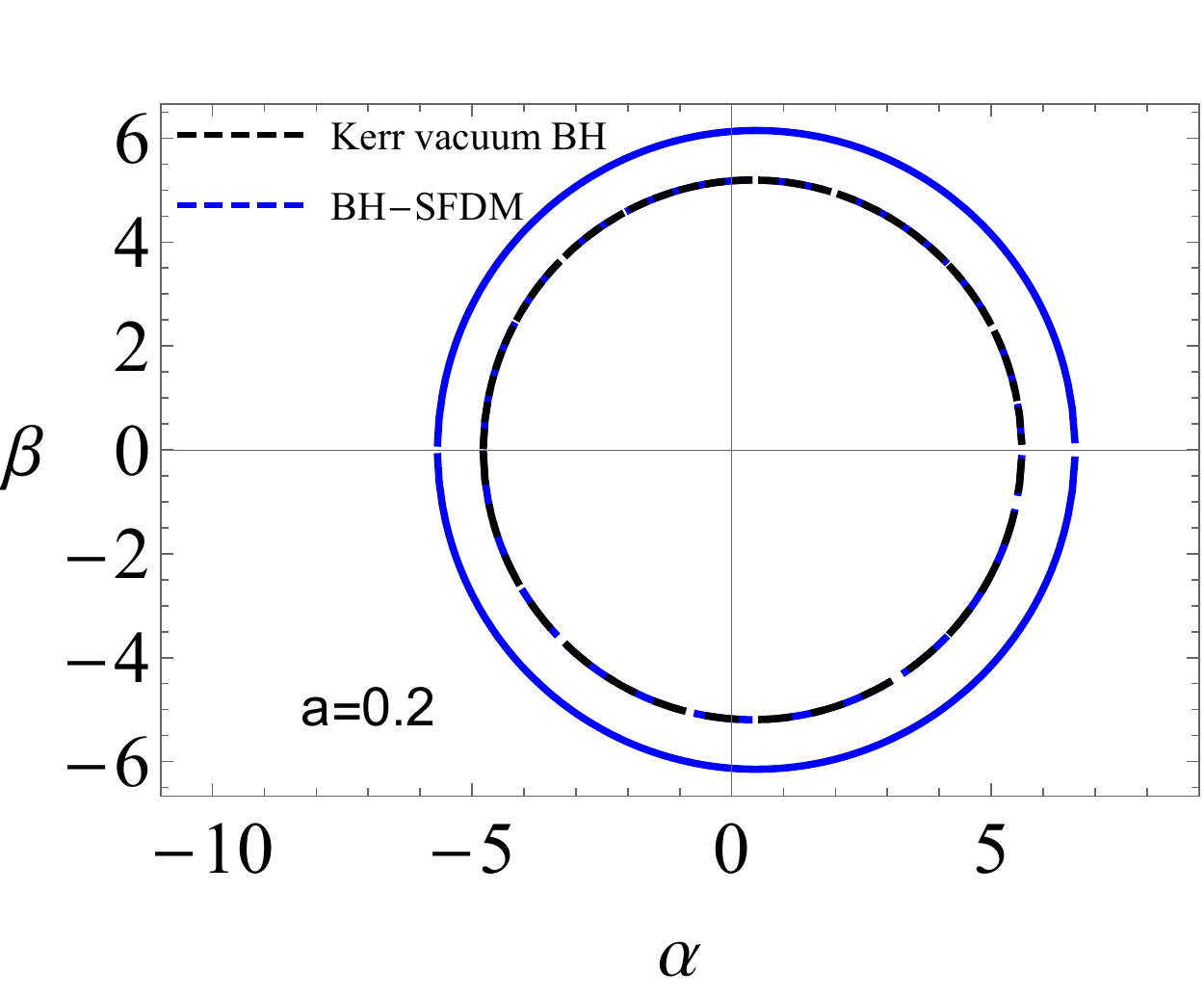}
\includegraphics[width=8.2cm]{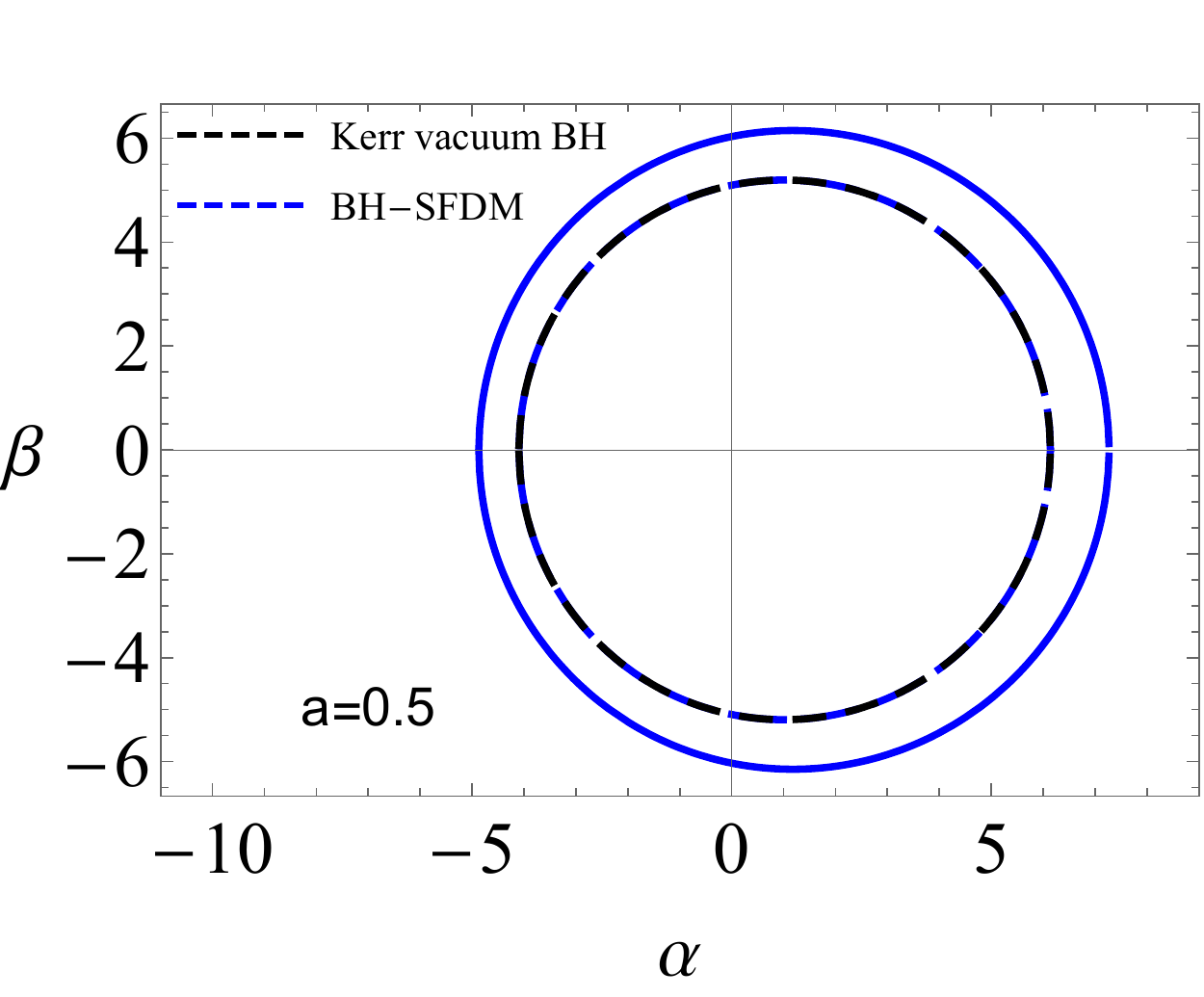}
\includegraphics[width=8.2cm]{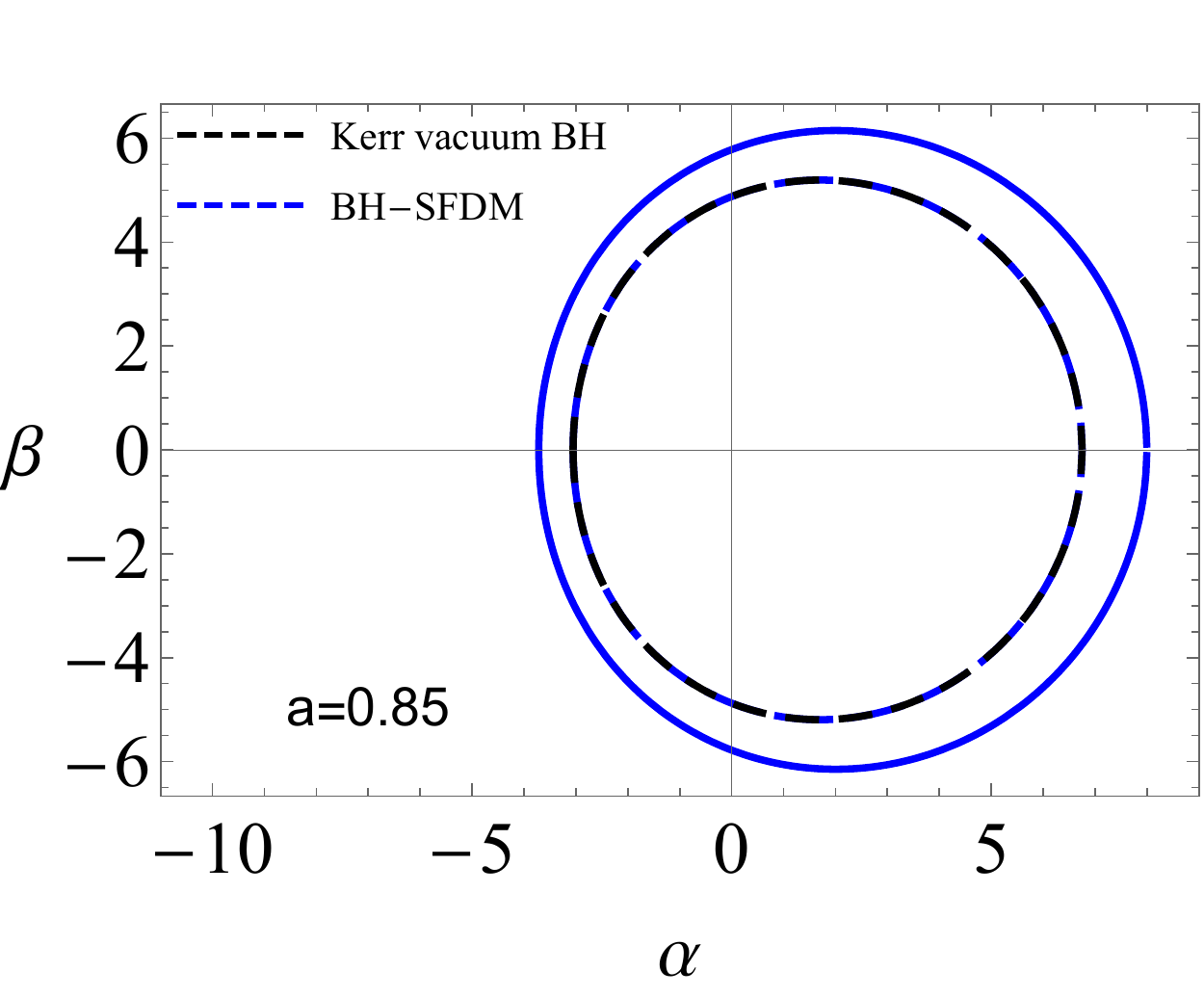}
\includegraphics[width=8.2cm]{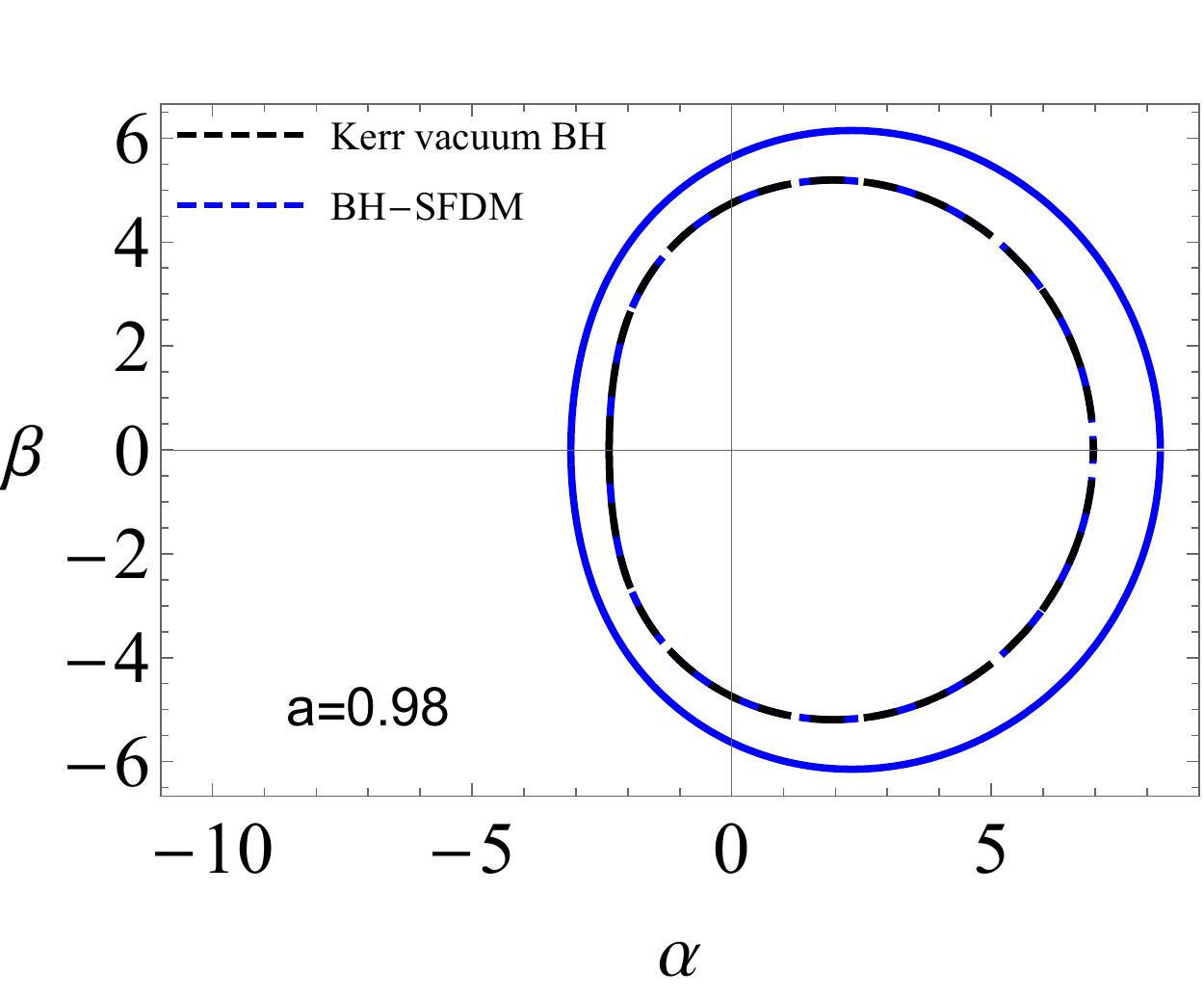}
\caption{Variation in shape of shadow using the spherical exponential profile for baryonic matter. We use $M=1$ in units of the Sgr A$^{*}$  black hole mass given by $M_{BH}=4.3\,\times 10^{6} M_{\odot}$ and $R=158$ kpc or $R=75.5\,\times 10^{10} M_{BH}$. Furthermore for $m=0.6$ eV and $\Lambda=0.2$ meV we find $\rho_0=0.02 \,\times 10^{-24}$ g/cm$^3$, or in eV units, $\rho_0=9.2\,\times 10^{-8} $ eV$^4$. For the Milky Way galaxy we can use $M_B=6 \times 10^{10} M{\odot}=1.39\,\times 10^{4} M_{BH}$ and $L=2.6$ kpc$=1.24\,\times 10^{10}  M_{BH}$. We observe that the dashed blue curve corresponding to Sgr A$^{*}$ with the above parameters is almost indistinguishable from the black dashed curve describing the Kerr vacuum BH. The solid blue curve corresponds to the case of increasing the baryonic mass by a factor of $ 10^2$ and decrease of core radius by a factor of $10^{3}$. } \label{B}
\end{figure*}

\begin{figure*}
\includegraphics[width=8.2cm]{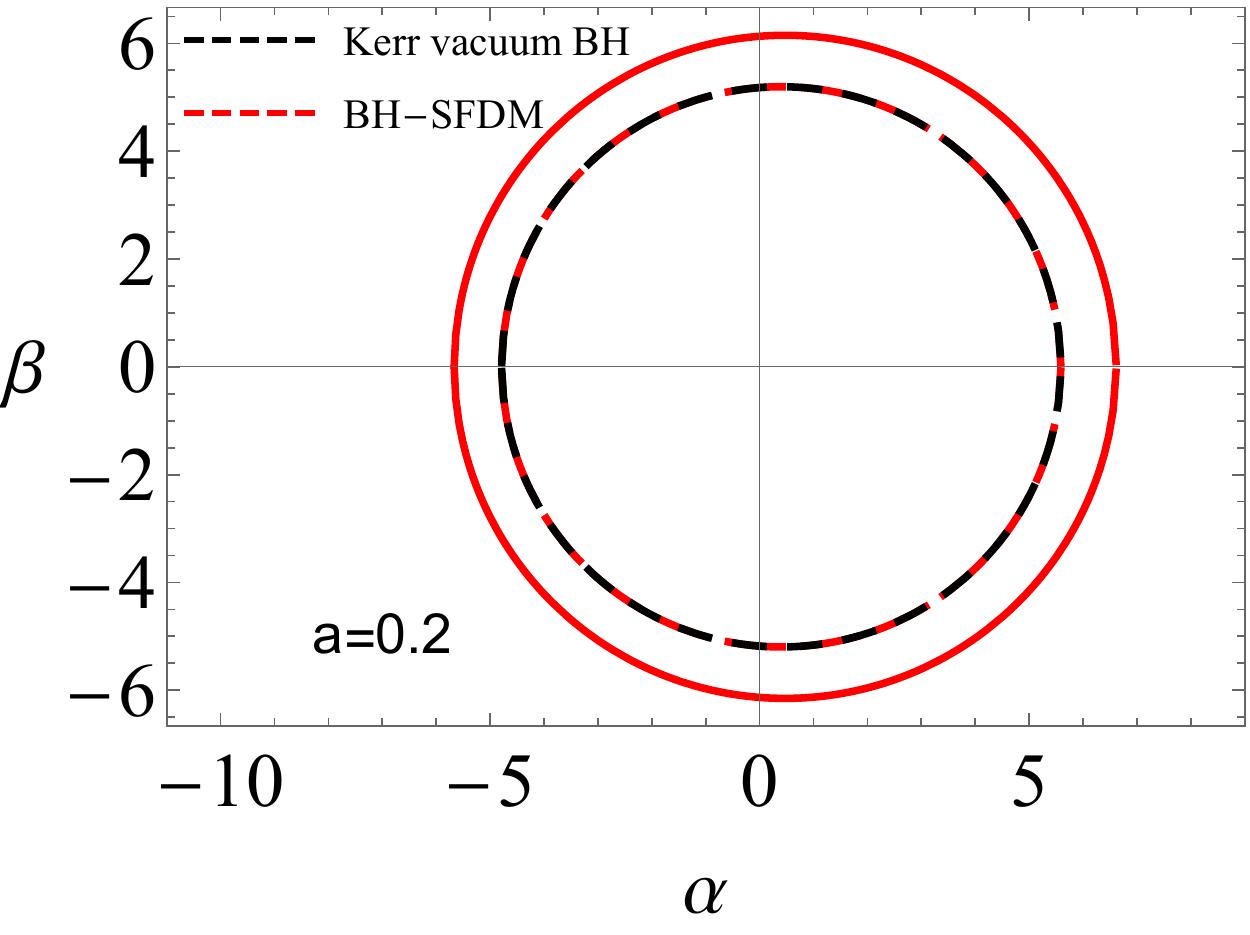}
\includegraphics[width=8.2cm]{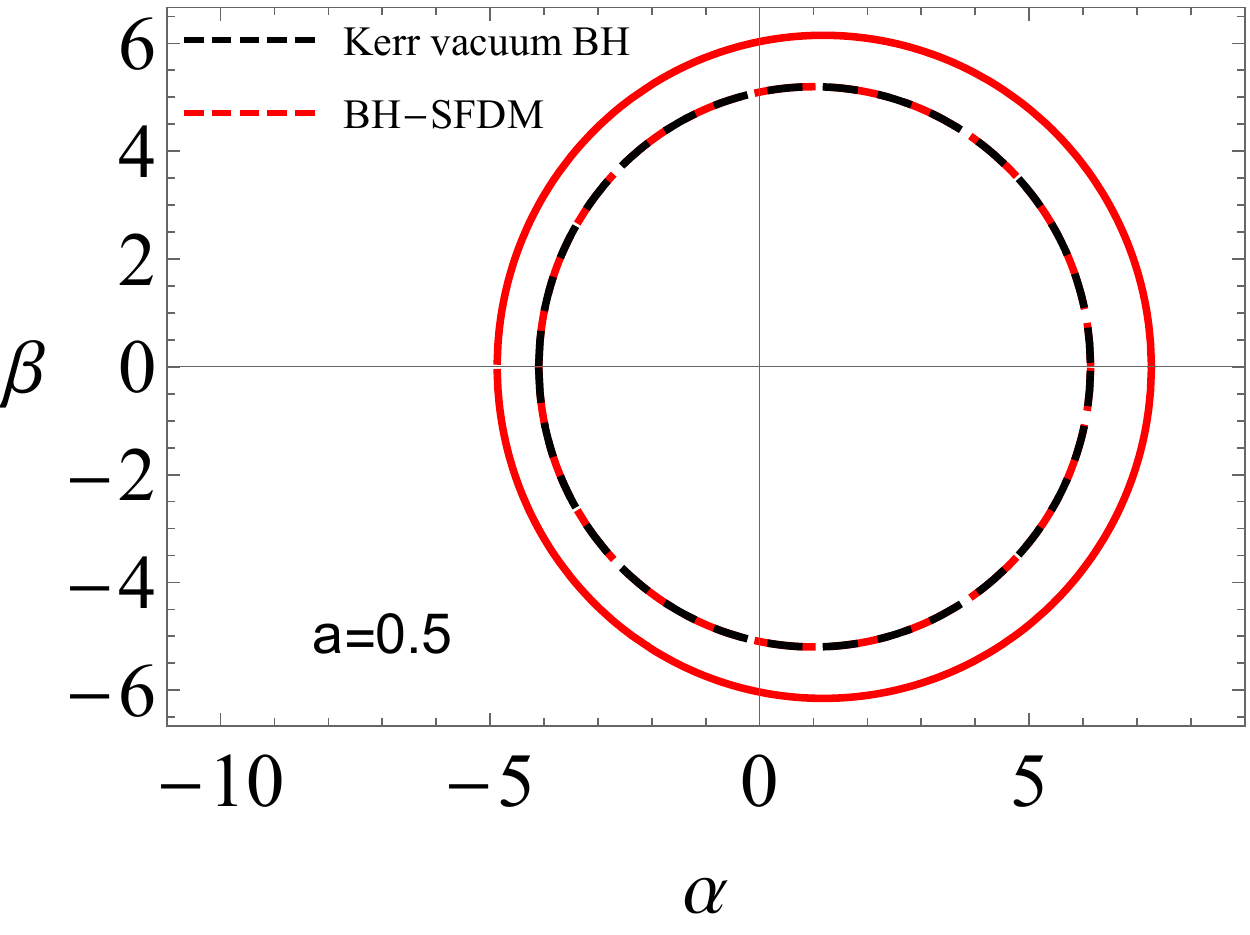}
\includegraphics[width=8.2cm]{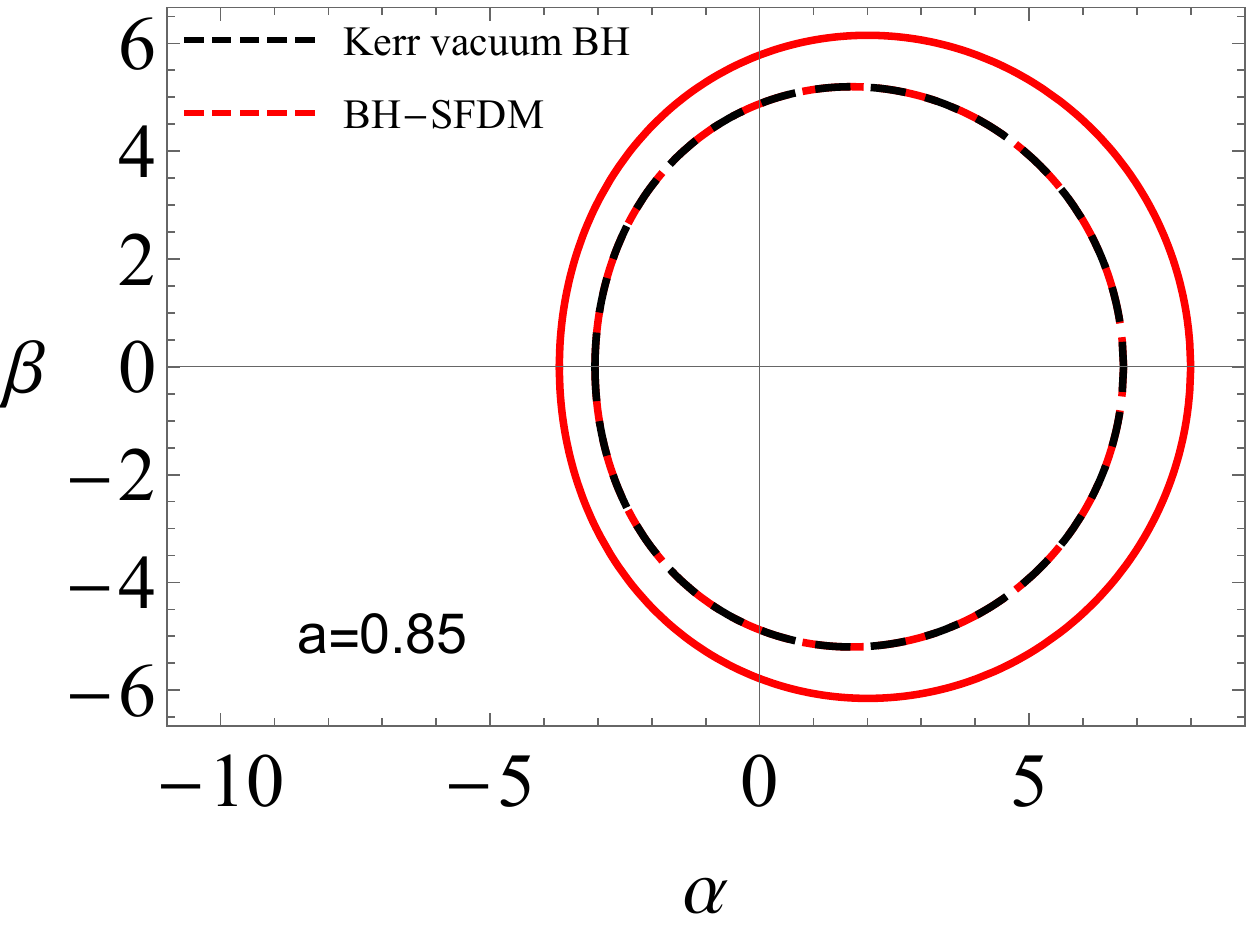}
\includegraphics[width=8.2cm]{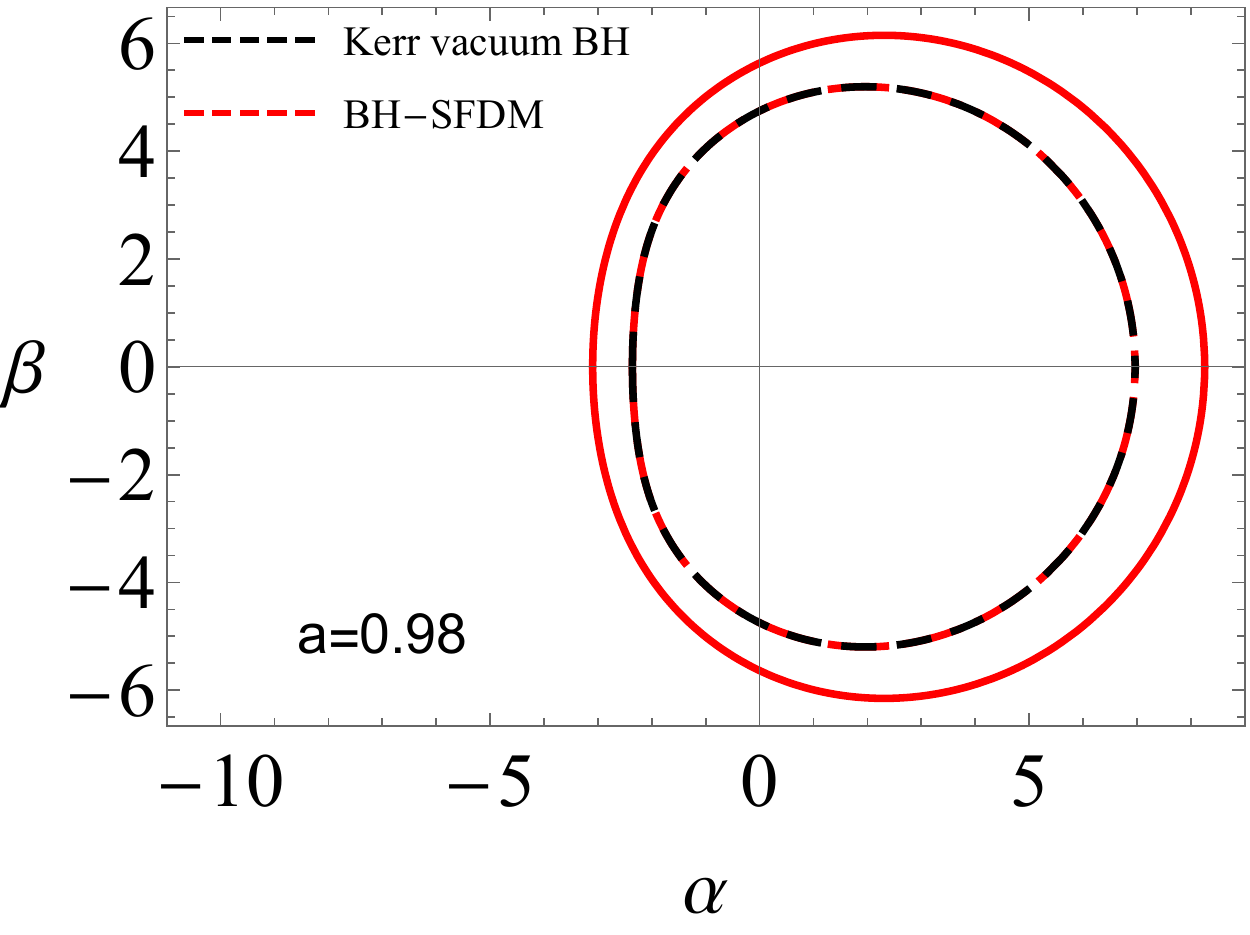}
\caption{Variation in shape of shadow for different values of $a$ using the power law profile. We use $M=1$ in units of the Sgr A$^{*}$  black hole mass given by $M_{BH}=4.3\,\times 10^{6} M_{\odot}$ and $R=158$ kpc or $R=75.5\,\times 10^{10} M_{BH}$. Furthermore for $m=0.6$ eV and $\Lambda=0.2$ meV we find $\rho_0=0.02 \,\times 10^{-24}$ g/cm$^3$, or in eV units, $\rho_0=9.2\,\times 10^{-8} $ eV$^4$. For the Milky Way galaxy we can use $M_B=6 \times 10^{10} M{\odot}=1.39\,\times 10^{4} M_{BH}$ and $r_c=2.6$ kpc$=1.24\,\times 10^{10}  M_{BH}$. The dashed red curve corresponds to Sgr A$^{*}$ using the above parameters which is almost indistinguishable from the black dashed curve discribing the Kerr vacuum BH. The solid red curve corresponds to the case of increasing the baryonic mass by a factor of $ 10^2$ and decrease of core radius by a factor of $10^{3}$.} \label{B}
\end{figure*}
\newpage
\subsection{Circular Orbits}
Due to the strong gravity near the black hole, it is thus expected that the photons emitted near a black hole will eventually fall into the black hole or eventually scatter away from it. In this way, the photons captured by the black hole will form a dark region defining the contour of the shadow. To elaborate the presence of unstable circular orbits around the black hole we need to study the radial geodesic by introducing the effective potential $V_{\text{eff}}$ which can be written as follows
\begin{equation}
\Sigma^2\left(\frac{dr}{d\tau}\right)^2+V_{\text{eff}}^{1,2}=0.
\end{equation}
where $V_{\text{eff}}^{1}=V_{\text{eff}}^{exp}$ and $V_{\text{eff}}^{2}=V_{\text{eff}}^{power}$, gives the exponential and the power law respectively. At this point, it is convenient to introduce two parameters $\xi$ and $\eta$, defined as
\begin{equation}
\xi=L/E,  \;\; \;\; \;\;   \eta=\mathcal{K}/E^2.
\end{equation}
For the effective potential then we obtain the following relation
\begin{equation}\label{veff}
V_{\text{eff}}^{1,2}=\Delta_{1,2}((a-\xi_{1,2})^2+\eta_{1,2})-(r^2+a^2-a\;\xi_{1,2})^2,
\end{equation}
where we have replaced $V^{1,2}_{\text{eff}}/E^2$ by $V^{1,2}_{\text{eff}}$. For more details in Figure (3) we plot the variation of the effective potential associated with the radial motion of photons. The circular photon orbits exists when at some  constant $r=r_{cir.}$ the conditions
\begin{equation}\label{cond}
V_{\text{eff}}^{1,2}(r)=0,\quad~~~\frac{dV^{1,2}_{\text{eff}}(r)}{dr}=0
\end{equation}

Combining all these equations it is possible to show that
\begin{equation}\notag
\xi_{1,2}= \frac{(r^2+a^2)(r \mathcal{F}'_{1,2}(r)+2\mathcal{F}_{1,2}(r))-4(r^2 \mathcal{F}_{1,2}(r)+a^2)}{a(r\mathcal{F}'_{1,2}(r)+2 \mathcal{F}_{1,2}(r))}
\end{equation}
\begin{equation}
\eta_{1,2}= \frac{r^3 [8 a^2 \mathcal{F}'_{1,2}(r)-r(r\mathcal{F}'_{1,2}(r)-2\mathcal{F}_{1,2}(r))^2]}{a^2(r\mathcal{F}'_{1,2}(r)+2\mathcal{F}_{1,2}(r))^2}.
\end{equation}

One can recover the Kerr vacuum case by letting $a_0=M_B=\rho_0= 0$, yielding
\begin{align}
\xi_{1,2}=\frac{r^2(3M-r)-a^2(M-r)}{a (r-M)},
\end{align}
\begin{align}
\eta_{1,2}=\frac{r^3\left(4 Ma^2 -r (r-3M)^2\right)}{a^2 (r-M)^2}.
\end{align}

To obtain the shadow images of our black hole in the presence of dark matter we assume that the observer is located at the position with coordinates $(r_o,\theta_o)$, where $r_o$ and $\theta_o$ represents the angular coordinate on observer's sky. Furthermore we need to introduce two celestial coordinates, $\alpha$ and $\beta$, for the observer by using the following relations \cite{Hioki:2008zw}
\begin{equation}
\alpha=-r_o \frac{p^{(\phi)}}{p^{(t)}},\,\,\,\,\beta=r_o\frac{p^{(\theta)}}{p^{(t)}},
\end{equation}
where $(p^{(t)},p^{(r)},p^{(\theta)},p^{(\phi)})$ are the tetrad components of
the photon momentum with respect to locally non-rotating reference frame. The observer bases $e^{\mu}_{(\nu)}$  can be expanded as a form in the coordinate bases (see, \cite{Cunha:2016bpi})
\begin{eqnarray}\notag
p^{(t)}&=&-e^{\mu}_{(t)}p_{\mu}=E\,\zeta_{1,2}-\gamma_{1,2}\, L,\,\, p^{(\phi)}=e^{\mu}_{(\phi)}p_{\mu}=\frac{L}{\sqrt{g_{\phi\phi }}},\\
p^{(\theta )}&=& e^{\mu}_{(\theta)}p_{\mu}=\frac{p_{\theta }}{\sqrt{g_{\theta \theta}}},\,\,p^{(r)}=e^{\mu}_{(r)}p_{\mu}=\frac{p_r}{\sqrt{g_{rr }}},
\end{eqnarray}
where the quantities $E =-p_t$ and $p_{\phi}=L$ are conserved due to the associated Killing vectors. If we use $\xi=L/E$, $\eta=\mathcal{K}/E^2 $ and $p_{\theta}=\pm \sqrt{\Theta(\theta)}$ we can rewrite these coordinates in terms our parameters $\xi$ and $\eta$, as follows  \cite{Kumar}
\begin{align}\nonumber
& \alpha_{1,2} = -r_o \dfrac{\xi_{1,2}}{\sqrt{g_{\phi \phi}}(\zeta_{1,2}-\gamma_{1,2} \xi_{1,2})}|_{(r_o,\theta_o)},\\\label{beta}
& \beta_{1,2} = \pm r_o \frac{\sqrt{\eta_{1,2}+a^2 \cos^2\theta-\xi_{1,2}^2 \cot^2\theta}}{\sqrt{g_{\theta \theta}}(\zeta_{1,2}-\gamma_{1,2} \xi_{1,2})}|_{(r_o,\theta_o)}.
\end{align}
where
\begin{equation}
    \zeta_{1,2}=\sqrt{\frac{g_{\phi\phi}}{g_{t \phi}^2-g_{tt}g_{\phi \phi}}}
\end{equation}
and
\begin{equation}
    \gamma_{1,2}=-\frac{g_{t \phi}}{g_{\phi \phi}}\zeta_{1,2}.
\end{equation}
Finally to simplify the problem further we are going to consider that our observer is located in the equatorial plane ($\theta=\pi/2$) and very large but finite $r_o=D=8.3$ kpc, we find
\begin{align}\nonumber
& \alpha_{1,2} = -\sqrt{f_{1,2}}\, \xi_{1,2},\\\label{beta}
& \beta_{1,2} = \pm \sqrt{f_{1,2}} \,\sqrt{\eta_{1,2}}.
\end{align}
Note that $f_{1,2}$ are given by Eqs. (32) and (34), respectively. We see that due to the presence of SFDM our solution is non-asymptotically flat. In the special case when SFDM is absent we recover $f_{1,2}\to 1$, hence the above relations reduces to the asymptotically flat case.   Note that for the observer located at the position with coordinates $(r_o,\theta_o)$ we have neglected the effect of rotation while $\xi_{1,2}$ and $\eta_{1,2}$ given by Eq. (61) and are evaluated at the circular photon orbits $r=r_{cir.}$
 \begin{figure}[h!]
    \includegraphics[width=0.45\textwidth]{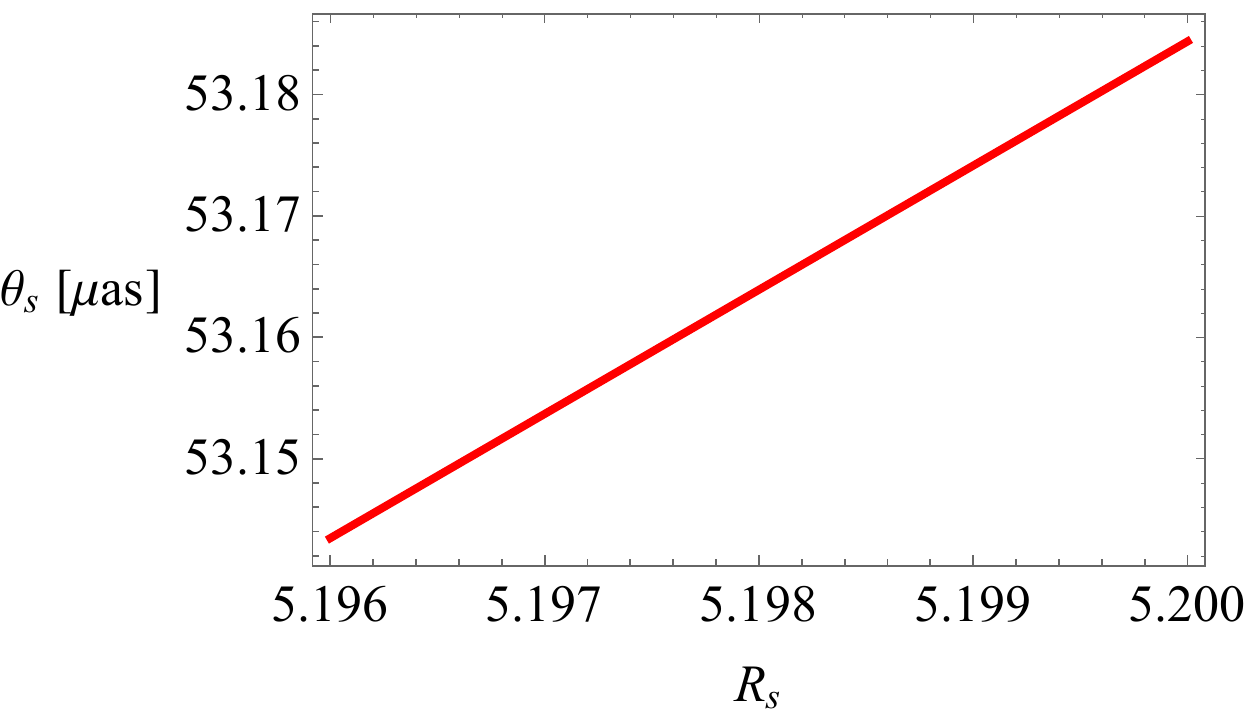}
  \caption{\label{figure1} The expected value for the angular diameter in the case of Sgr A$^{*}$ . Note that we have adopted $D =8.3$ kpc and $M = 4.3 \times  10^{6}$M\textsubscript{\(\odot\)}.  }
  \end{figure}

For more details in Figs. 5 and 6 we plot the shape of Sgr A$^{*}$ black hole shadows. Using the parameter values specifying the SFDM and baryonic matter we obtain almost no effect on the shadow images. In fact one can observe that the dashed curve (exponential profile) and dashed red curve (power law profile) are almost indistinguishable from the black dashed curve (Kerr vacuum BH).  However we can observe that by increasing the baryonic mass the shadow size increases considerably (solid blue/red curve). The angular radius of the Sgr A$^{*}$ black hole shadow can be estimated using the observable $R_s$ as $\theta_s = R_s M/D$, where $M$ is the black hole mass and $D$ is the distance between the black hole and the observer. The angular radius can be further expressed
as $\theta_s = 9.87098 \times 10^{-6} R_s(M/$M\textsubscript{\(\odot\)})$(1kpc/ D)$ $\mu$as. In the case of Sgr A$^{*}$,  we have used $M = 4.3 \times  10^{6}$M\textsubscript{\(\odot\)} and $D =8.3$ kpc \cite{Hou:2018bar} is the distance between the Earth and Sgr A$^{*}$ center black hole.  
In order to get more information about the effect of SFDM on physical observables we shall estimate the effect of SFDM on the angular radius for Sgr A$^{*}$.  To simplify the problem, let us consider our static and spherically symmetric black hole metric in a totally dominated dark matter galaxy described by the function
\begin{equation}
\mathcal{F}(r)=e^{-X}-\frac{2M}{r},
\end{equation}
where we have introduced
\begin{equation}
X=\frac{\rho_0^2 }{4 \Lambda^2 m^6 },
\end{equation}
where we have used the approximation $\cos(\frac{\pi \,r}{2 \,R}) \simeq 1$ since our solution is valid for $r \leq R$. From the circular photon orbit conditions one can show \cite{Zhu:2019ura}
\begin{equation}\lb{PSradius}
2 - \frac{r \mathcal{F}'(r)}{\mathcal{F}(r)}=0.
\end{equation}
Now if we expand around $X$ in the above function we obtain
\begin{equation}
\mathcal{F}(r)=1-X-\frac{2M}{r}+...
\end{equation}
By solving this equation and considering only the leading order terms one can determine the radius of the photon sphere $r_{\rm ps}$ yielding
\begin{equation}
r_{\rm ps}=3M(1+X).
\end{equation}
On the other hand, one also can show the relation \cite{Zhu:2019ura}
\begin{equation}\label{eq74}
\xi^2 +\eta = \frac{r^2_{\rm ps}}{ \mathcal{F}(r_{\rm ps})}. 
\end{equation}
The shadow radius $R_{\rm s}$ can be expressed in terms of the celestial coordinates $(\alpha,\beta)$ as follows
\begin{equation}\lb{RS}
R_{\rm s} = \sqrt{\alpha^2+\beta^2} =\sqrt{1-X}\frac{r_{\rm ps}}{ \sqrt{\mathcal{F}(r_{\rm ps})}}.
\end{equation}
Note that in the last equation we have used Eqs. \eqref{beta} and \eqref{eq74}. 
Using the values:  $m=0.6$ eV, $\Lambda=0.2$ meV and $\rho_0=9.2\,\times 10^{-8} $ eV$^4$, in units of the Sgr A$^{*}$ black hole we find $R_{\rm s}=5.196158313 M$. Using this result, we can find that the angular diameter increases by $\delta \theta_s= 3 \times 10^{-5}$ $\mu$as compared to the Schwarzschild vacuum. This result is consistent with the result reported in Ref. \cite{Hou:2018bar} where authors estimated that the dark matter halo could influence the shadow of Sgr A$^{*}$ at a level of order of magnitude of $10^{-3}$ $\mu$as and $10^{-5}$ $\mu$as, respectively. They have used the so called Cold Dark Matter and Scalar Field Dark Matter models to study the apparent shapes of the shadow. A similar result has been obtained for the dark matter effect on the M87 black hole (see, \cite{Jusufi:2019nrn}) using the Burkert dark matter profile.

\section{Conclusion}

In this paper we have obtained a rotating black hole solution surrounded by superfluid dark matter along with baryonic matter.  To achieve this purpose, in the present work, we considered the superfluid dark matter model and used two specific profiles to describe the baryonic matter distribution, namely the spherical exponential profile and the power law profile followed by the special case of a totally dominated dark matter case. Using the current values for the baryonic mass, central density of the superfluid dark matter, halo radius, as well as the radial scale length for the baryonic matter in our galaxy, we found that the shadow size of Sgr A$^{*}$ black hole remains almost unchanged compared to the Kerr vacuum BH. This result is consistent with a recent work reported in  \cite{Jusufi:2019nrn} and also \cite{Hou:2018bar}.  For entirely dominated dark matter galaxies, we find that the angular diameter increases by $10^{-5} \mu$arcsec. This result shows that it is very difficult to constrain the dark matter parameters using the shadow images. Such tiny effects on the angular diameter are out of reach for the existing space technology and it remains an open question if future astronomical observations can potentially detect such effects. That being said, the expected value for the angular diameter in the case of Sgr A$^{*}$ is of the order of $\theta_s \simeq 53\, \mu$as as predicted in GR. 

As an interesting observation, we show that an increase of baryonic mass followed by a decrease of the radial scale length can increase the shadow size considerably. This can be explained by the fact that the baryonic matter is mostly located in the interior of the galaxy, on the other hand, dark matter is mostly located in the outer region of the galaxy. In other words, for the totally dominated dark matter galaxies we observe almost no effect on black hole shadows, but a more precise measurement of the Sgr A$^{*}$ shadow radius can play a significant role in determining the baryonic mass in our galaxy.

\section*{Acknowledgements}
TZ is supported by National Natural Science Foundation of China under the Grants No. 11675143, the Zhejiang Provincial Natural Science Foundation of China under Grant No. LY20A050002, and the Fundamental Research Funds for the Provincial Universities of Zhejiang in China under Grants No. RF-A2019015.

\appendix
\section{}\label{Apend_a}
Here we demonstrate the derivation of the spinning black hole metric. As a first step to this formalism, we transform Boyer-Lindquist (BL) coordinates $(t,r,\theta,\phi)$ to Eddington-Finkelstein (EF) coordinates $(u,r,\theta,\phi)$. This can be achieved by using the coordinate transformation
\begin{eqnarray}\label{eq18}
dt&=&du+\frac{dr}{\sqrt{\mathcal{F}_{1,2}(r)\mathcal{G}_{1,2}(r)}},
\end{eqnarray}
we obtain
\begin{equation}
ds^2_{1,2}=-\mathcal{F}_{1,2}(r) du^2-2 \sqrt{\frac{\mathcal{F}_{1,2}(r)}{\mathcal{G}_{1,2}(r)}}du dr+r^2 d\theta^2+r^2 \sin^2\theta d\phi^2.
\end{equation}

This metric can be expressed in terms of null tetrads as
\begin{eqnarray}
g^{\mu{\nu}}=-l^{\mu}n^{\nu}-l^{\nu}n^{\mu}+m^{\mu}\overline{m}^{\nu}+m^{\nu}\overline{m}^{\mu},
\end{eqnarray}
where the null tetrads are defined as
\begin{eqnarray}
l^{\mu}&=&\delta^{\mu}_{r},\\
n^{\mu}&=& \sqrt{\frac{\mathcal{G}_{1,2}(r)}{\mathcal{F}_{1,2}(r)}} \delta^{\mu}_{u}-\frac{1}{2}\mathcal{G}_{1,2}(r)\delta^{\mu}_{r},\\
m^\mu&=&\frac{1}{\sqrt{2\mathcal{H}}}\left(\delta^{\mu}_{\theta}+\frac{\dot{\iota}}{\sin\theta}\delta^{\mu}_{\phi}\right).
\end{eqnarray}
These null tetrads are constructed in such a way that $l^\mu$ and $n^\mu$ while  $m^\mu$ and $\bar{m}^\mu$ are complex. It is worth noting that $\mathcal{H}=r^2$ and will be used later on. It is obvious from the notation that $\bar{m}^\mu$ is complex conjugate of $m^\mu$. These vectors further satisfy the conditions for normalization, orthogonality and isotropy as
\begin{eqnarray}
l^{\mu}l_{\mu}=n^{\mu}n_{\mu}=m^{\mu}m_{\mu}=\bar{m}^{\mu}\bar{m}_{\mu}=0,\\
l^{\mu}m_{\mu}=l^{\mu}\bar{m}_{\mu}=n^{\mu}m_{\mu}=n^{\mu}\bar{m}_{\mu}=0,\\
-l^{\mu}n_{\mu}=m^{\mu}\bar{m}_{\mu}=1.
\end{eqnarray}
Following the Newman--Janis prescription we write,
\begin{equation}
{x'}^{\mu} = x^{\mu} + ia (\delta_r^{\mu} - \delta_u^{\mu})
\cos\theta \rightarrow \\ \left\{\begin{array}{ll}
u' = u - ia\cos\theta, \\
r' = r + ia\cos\theta, \\
\theta' = \theta, \\
\phi' = \phi. \end{array}\right.
\end{equation}
in which $a$ stands for the rotation parameter. Next, let the null tetrad vectors $Z^a$ undergo a
transformation given by $Z^\mu = ({\partial x^\mu}/{\partial {x^\prime}^\nu}) {Z^\prime}^\nu $, following \cite{Azreg-Ainou:2014pra}
\begin{eqnarray}
l'^{\mu}&=&\delta^{\mu}_{r},\\\notag
n'^{\mu}&=&\sqrt{\frac{B_{1,2}}{A_{1,2}}}\delta^{\mu}_{u}-\frac{1}{2}B_{1,2}\delta^{\mu}_{r},\\\notag\label{e11}
m'^{\mu}&=&\frac{1}{\sqrt{2\Sigma_{1,2}}}\left[(\delta^{\mu}_{u}-\delta^{\mu}_{r})\dot{\iota}{a}\sin\theta+\delta^{\mu}_{\theta}+\frac{\dot{\iota}}{\sin\theta}\delta^{\mu}_{\phi}\right]
\end{eqnarray}
where  we have assumed that $(\mathcal{F}_{1,2}(r),\mathcal{G}_{1,2}(r), \mathcal{H}(r))$ transform to $(A_{1,2}(a,r,\theta), B_{1,2}(a,r,\theta), \Sigma_{1,2}(a,r,\theta))$. With the help of the above equations new metric in Eddington-Finkelste coordinate reads
\begin{eqnarray}
ds_{1,2}^2&=&-A_{1,2}du^2-2\sqrt{\frac{A_{1,2}}{B_{1,2}}}dudr\\\notag
&+&2a\sin^2\theta\left(A_{1,2}-\sqrt{\frac{A_{1,2}}{B_{1,2}}}\right)du d\phi\\\notag
&+& 2a\sqrt{\frac{A_{1,2}}{B_{1,2}}}\sin^2\theta drd\phi+\Sigma_{1,2} d\theta^2 \\\notag
&+&\sin^2\theta\left[\Sigma_{1,2}+a^2\sin^2\theta\left(2\sqrt{\frac{A_{1,2}}{B_{1,2}}}-A_{1,2}\right)\right]d\phi^2.
\end{eqnarray}

Note that $A_{1,2}$  is some function of $r$ and $\theta$ as we already pointed out. Without going into details of the calculation one can revert the EF coordinates back to BL coordinates by using the following transformation (see for more details \cite{Azreg-Ainou:2014pra})
\begin{eqnarray}
du=dt- \frac{a^2+r^2}{\Delta_{1,2}}dr,\,\,\,\,\, d\phi=d\phi'- \frac{a}{\Delta_{1,2}}dr,
\end{eqnarray}
where in order to simplify the notation we introduce 
\begin{equation}
\Delta(r)_{1,2}=r^2\,\mathcal{F}_{1,2}(r)+a^2.
\end{equation}

Making use of $\mathcal{F}_{1,2}(r)=\mathcal{G}_{1,2}(r)$, one can obtain $\Sigma_{1}=\Sigma_{2}=r^2+a^2\cos^2\theta$, herein we shall use just $\Sigma$. 
Dropping the primes in the $\phi$ coordinate and choosing 
\begin{equation}
A_{1,2}=\frac{(\mathcal{F}_{1,2}\mathcal{H}+a^2 \cos^2\theta)\Sigma_{1,2}}{(\mathcal{H}+a^2 \cos^2\theta)^2},
\end{equation}
and 
\begin{equation}
B_{1,2}=\frac{\mathcal{F}_{1,2}\mathcal{H}+a^2 \cos^2\theta}{\Sigma},
\end{equation}
we obtain the spinning black hole space-time metric in a SFDM halo 
\begin{eqnarray}\notag
ds^2_{1,2}&=&-\left(1-\frac{2\Upsilon_{1,2}(r) r}{\Sigma}\right)dt^2+\frac{\Sigma}{\Delta_{1,2}}dr^2+\Sigma d\theta^2 \\\notag
&-& 2 a \sin^2\theta \frac{2\Upsilon_{1,2}(r) r}{\Sigma}dt d\phi \\
& +&\sin^2\theta \left[\frac{(r^2+a^2)^2-a^2\Delta_{1,2} \sin^2\theta}{\Sigma} \right] d\phi^2
\end{eqnarray}
with
 \begin{equation}
\Upsilon_{1,2}(r)=\frac{r\, (1-\mathcal{F}_{1,2}(r))}{2},
\end{equation}
in which $\mathcal{F}_{1}=\mathcal{F}(r)_{exp}$ and $\mathcal{F}_{2}=\mathcal{F}(r)_{power}$, respectively. \\

\end{document}